\documentclass[screen,review=false, anonymous=false, sigconf]{acmart}

\AtBeginDocument{%
  \providecommand\BibTeX{{%
    \normalfont B\kern-0.5em{\scshape i\kern-0.25em b}\kern-0.8em\TeX}}}

\copyrightyear{2022}
\acmYear{2022}
\setcopyright{rightsretained}
\acmConference[AIES'22]{Proceedings of the 2022 AAAI/ACM Conference on AI, Ethics, and Society}{August 1--3, 2022}{Oxford, United Kingdom}
\acmBooktitle{Proceedings of the 2022 AAAI/ACM Conference on AI, Ethics, and Society (AIES'22), August 1--3, 2022, Oxford, United Kingdom}\acmDOI{10.1145/3514094.3534173}
\acmISBN{978-1-4503-9247-1/22/08}

\usepackage{subcaption}

\newcommand{\xhdr}[1]{\vspace{1mm} \noindent{\bf #1}.}

\setcopyright{none}
\renewcommand\footnotetextcopyrightpermission[1]{} %
\begin{document}

\title{Long-term Dynamics of Fairness Intervention in Connection Recommender Systems}

\author{Nil-Jana Akpinar}
\email{nakpinar@andrew.cmu.edu}
\affiliation{%
  \institution{Carnegie Mellon University}
  \streetaddress{5000 Forbes Avenue}
  \city{Pittsburgh}
  \state{Pennsylvania}
  \country{USA}
  \postcode{15213}
}

\author{Cyrus DiCiccio}  
\affiliation{%
  \institution{Work done while at LinkedIn}
  \streetaddress{1000 W Maude Ave}
  \city{Sunnyvale}
  \state{California}
  \country{USA}
  \postcode{94085}}
  
\author{Preetam Nandy}
\affiliation{%
  \institution{LinkedIn}
  \streetaddress{1000 W Maude Ave}
  \city{Sunnyvale}
  \state{California}
  \country{USA}
  \postcode{94085}}
  
\author{Kinjal Basu}
\affiliation{%
  \institution{LinkedIn}
  \streetaddress{1000 W Maude Ave}
  \city{Sunnyvale}
  \state{California}
  \country{USA}
  \postcode{94085}}

\renewcommand{\shortauthors}{Akpinar et al.}

\begin{abstract}
Recommender system fairness has been studied from the perspectives of a variety of stakeholders including content producers, the content itself and recipients of recommendations. 
Regardless of which type of stakeholders are considered, most works in this area assess the efficacy of fairness intervention by evaluating a single fixed fairness criterion through the lens of a one-shot, static setting. Yet recommender systems constitute dynamical systems with feedback loops from the recommendations to the underlying population distributions which could lead to unforeseen and adverse consequences if not taken into account.
In this paper, we study a connection recommender system patterned after the systems employed by web-scale social networks and analyze the long-term effects of intervening on fairness in the recommendations.
We find that, although seemingly fair in aggregate, common exposure and utility parity interventions fail to mitigate amplification of biases in the long term.
We theoretically characterize how certain fairness interventions impact the bias amplification dynamics in a stylized P\'{o}lya urn model.
\end{abstract}

\maketitle
\pagestyle{plain}

\section{Introduction}

Machine learning based recommender systems are at the heart of user experience in many social media applications. These systems underpin a wide range of services, including content ranking, connection recommendation, and job search tools. It is imperative that people participating in these systems, either as the recipient of ranked suggestions, or as the originator of the content being recommended are treated fairly which motivates a rich body of research in ranking and recommendation fairness \cite[e.g.][]{Zehlike_Castillo_2020,Yang_Stoyanovich_2016,Singh_Joachims_2019,Zehlike_Bonchi_Castillo_Hajian_Megahed_Baeza-Yates_2017,Celis_Straszak_Vishnoi_2017}.

Consider a connection recommendation setting where the system suggests a list of users based on a prompt such as `People you may know' and the recipient of the recommendation decides which of the users to connect with. How should the platform promote fairness between different user groups in these recommendations? 
An array of definitions, fairness enhancing algorithms and evaluation metrics have been proposed to address this and similar problems. %
Most approaches assume static prediction settings and focus on a single fairness metric in individual instances of the recommendation in a one-shot or time-aggregate manner. %
However, recommender systems are dynamic in nature with recommendations influencing user behavior and experience through time. This is particularly evident in people connection recommendation systems which lead to a connection graph that evolves over time. Limiting ourselves to only one targeted fairness metric while neglecting other important variables of interest and potential dynamics influenced by the intervention may lead to unintended consequences and overlooked side effects \cite{Dai_Fazelpour_Lipton_2021,damour2020}.

This work focuses on the long-term effects of fairness intervention in connection recommendation.
We empirically demonstrate that these systems can suffer from a group-wise `rich-get-richer' phenomenon which exacerbates outcome disparities over time. Through a simulation framework, we study the long-term impact of fairness interventions, finding that, although seemingly fair in aggregate, a key desiderata of fairness intervention, i.e. equity in network sizes, is not promoted over time. In fact, average network sizes diverge in the long run even with popular fairness interventions leaving the bulk of the minority group disadvantaged while simultaneously creating an illusion of fairness.
We support our empirical findings by conducting a theoretical limit analysis of the impact of different fairness interventions on bias amplification dynamics assuming a stylized connection recommendation system based on P\'{o}lya urns.

Understanding the potential long-term harms of mitigation approaches through online experimentation can be time consuming
and potentially lead to real harm.
A simulation framework allows us to gain insight into the impacts of various notions of fairness, and to better understand how applying fairness interventions translates into tangible outcomes. Also, because of issues of network interference, it can be challenging to fully understand the impacts of fairness interventions on a network graph through experimentation. For these reasons, we defer to simulation along with theory that supports the simulation findings.
Simulation-driven methods to uncover long-term effects have been previously used in the recommendation literature following the observation that offline experiments on observational data are often insufficient to assess performance after deployment \cite{Krauth2020,GomezUribe2016}, and they provide a promising path towards understanding context specific fairness dynamics in the ranking and recommendation fairness setting \cite{patro2022fair}.

The remainder of the paper is organized as follows. Section \ref{sec:related} provides an overview of related fairness literature including fairness in recommender systems, mitigation approaches, and long-term dynamics.
Section \ref{sec:methodology} outlines fairness criteria, corresponding methodology for mitigation, and a simulator modelling a connection recommender system. Results of the simulation study are given in Section \ref{sec:emp_results}. Section \ref{sec:theory} derives theoretical results demonstrating the validity of the empirical results, and leverages this theory to understand the workings of the fairness interventions. Finally, we conclude with a discussion in Section \ref{sec:discussion}.

\section{Background and related work}\label{sec:related}

\subsection{Fairness in recommendations}

The most well-studied types of recommender system bias include popularity bias which refers to the over-recommendation of already popular items \cite[e.g.][]{Jannach_Lerche_Kamehkhosh_Jugovac_2015,Abdollahpouri_Mansoury_2020,abdollahpouri2019unfairness}, and position bias which describes the tendency of members to interact primarily with top ranked items when shown recommendations in the form of a list \cite[e.g.][]{Joachims2007,Joachims2017,Craswell2008,Wang2018}. 
More recently, researchers have started to take interest in outcome disparities on a group level which can have an intricate relationship with known deficiencies like popularity bias \cite{Ekstrand_Tian_Azpiazu_Ekstrand_Anuyah_McNeill_Pera_2018,abdollahpouri2019unfairness}.
While some work considers group level disparities on the source side, i.e. for the members who implicitly query and receive the recommendations \cite[e.g.][]{Ekstrand_Tian_Azpiazu_Ekstrand_Anuyah_McNeill_Pera_2018}, a considerable body of research concentrates on fairness for the destination side, i.e. the items or people being recommended. 
This focus is based on the common understanding that exposure in recommendations is a valuable but scarce resource that can be the deciding factor in which suppliers can sell their items or who gets a job offer.

A variety of different metrics and bias mitigation algorithms for fairness in recommendation lists have been proposed \cite[e.g.][]{Mehrotra_McInerney_Bouchard_Lalmas_Diaz_2018,Zehlike_Bonchi_Castillo_Hajian_Megahed_Baeza-Yates_2017,Yang_Stoyanovich_2016,Celis_Straszak_Vishnoi_2017,Beutel_Chen_Doshi_Qian_Wei_Wu_Heldt_Zhao_Hong_Chi_et,Nandy_Diciccio_Venugopalan_Logan_Basu_El}.
For example, \cite{Mehrotra_McInerney_Bouchard_Lalmas_Diaz_2018} use counterfactual estimation techniques to understand the impact of different recommendation policies in the context of music recommendation which has been shown to suffer from gender bias \cite{Ferraro_Serra_Bauer_2021,shakespeare2020exploring}.
A different line of work relies on pairwise comparisons from randomized experiments to measure fairness in rankings \cite{Beutel_Chen_Doshi_Qian_Wei_Wu_Heldt_Zhao_Hong_Chi_et}. %
The works of \cite{Zehlike_Bonchi_Castillo_Hajian_Megahed_Baeza-Yates_2017} and \cite{Yang_Stoyanovich_2016} are concerned with the problem of top-$k$ ranking, and 
the authors of \cite{Nandy_Diciccio_Venugopalan_Logan_Basu_El} propose post-processing methods to achieve equality of opportunity or equalized odds in recommender systems.
A straightforward but flexible statistical notion of fairness in the context of recommendations is demographic parity of exposure which has been used by a number of papers.
\cite{Singh_Joachims_2018} maximize ranking utility for the viewer of a ranked list subject to exposure-centered fairness constraints including demographic parity.
The work of \cite{Zehlike_Castillo_2020} uses an in-processing approach that directly focuses on enforcing demographic parity of exposure in scoring models used for ranking.
\cite{Abdollahpouri_Adomavicius_Burke_Guy_Jannach_Kamishima_Krasnodebski_Pizzato_2019} analyze how popularity bias affects different stakeholders of recommender systems and propose exposure-based metrics such as demographic parity to measure unfairness.

Based on the popularity of the metric, our analysis of long-term dynamics of fairness interventions in connection recommender systems begins by assuming demographic parity of exposure as fairness measure.
The empirical portion of this work assumes a probabilistic ranking framework similar to the settings in \cite{Basu_DiCiccio_Logan_El,Singh_Joachims_2018}, and tracks the effects of demographic parity of exposure and other fairness metrics in connection recommendation over an extended period of time which enables us to make observations that have previously been overlooked.

\subsection{Feedback loops and long-term fairness}

Evaluation of fairness in recommender systems often focuses on single recommendation steps or time-aggregate behavior of the system.
An exception to this is a body of work on popularity bias which has been shown to be highly dynamic over time
\cite{masoud_twitter,Yao_Halpern_Thain_Wang_Lee_Prost_Chi_Chen_Beutel_2021,Chen_Dong_Wang_Feng_Wang_He_2020}.
In many applications, popularity bias leads to a feedback loop that further increases the exposure of popular items over time leading to a long-tail phenomenon often called ``rich-get-richer effect'' or ``Matthew effect''.
\cite{Chen_Dong_Wang_Feng_Wang_He_2020} summarize methods proposed to break this feedback loop including reliance on uniform data \cite{Jiang2019,Liu2020} and reinforcement learning-based recommenders which are able to adapt to changing states of the system \cite{Zhao2019,Ge_Liu_Gao_Xian_Li_Zhao_Pei_Sun_Ge_Ou_et}.
Both of these approaches are challenging to realize in real-world applications.
While obtaining uniform data in practice generally requires deploying some sort of uniformly at random recommendation policy which hurts member experience \cite{Chen_Dong_Wang_Feng_Wang_He_2020}, reinforcement learning-based recommenders are difficult to evaluate since they only have access to data biased by an existing policy \cite{Jagerman2019,Chen2019_offpolicy}.
To the best of our knowledge, few works have considered how intervening on fairness between different demographic groups may change the dynamics of a recommender system in the long term.
\cite{Ferraro_Serra_Bauer_2021} study the setting of music recommendation and observe a positive feedback loop when intervening on gender fairness. The authors propose an iterative reranking approach to mitigate unfairness measured based on a number of task-specific metrics.
Simulation results suggest that, over time, intervention leads the music recommendation algorithm to make fairer recommendations organically.
\cite{Morik_Singh_Hong_Joachims_2020} incorporate a potential feedback loop into their intervention procedure, and propose an algorithm in the form of a controller that optimizes utility under amortized group fairness constraints while dynamically adapting as more data becomes available.

In this work, we study how the addition of common statistical notions of fairness to probabilistic ranking problems for connection recommendation impacts the state of a social network in the long term. Our variable of interest is the difference in average network sizes between groups which can be understood as a measure of diverging benefits from the recommender system.
This follows a similar idea as \cite{Liu_Dean_Rolf_Simchowitz_Hardt_2018} who, while not directly concerned with recommender systems, study the delayed impact of fairness intervention in a classification setting with a one-step feedback loop. The authors assume a hypothetical lending scenario in which lending decisions are based on and impact the score distributions in two demographic groups. Here, the central quantity of interest is the difference between average scores in groups and the findings suggest that common fairness criteria do not necessarily promote improvement over time.

\section{Methodology}\label{sec:methodology}

\subsection{Optimization framework for connection recommendation}

Connection recommender systems generally rely on a member's current network as well as other member data in order to select a group of members of the platform to suggest as potential connections. 
Expanding one's network is assumed to create positive value for the both the member at the source side (viewer of the recommendation) and the member at the destination side (recommended member).
Recommendations generally do not require specific search terms or prompts by the source member but are instead often provided in a single or a few automatically generated categories, e.g. `People you may know' or `People you may know from your workplace'. 
Each member of the platform simultaneously serves as a source and a destination member and is thus associated to a two dimensional vector of recommendation utilities.
Despite stakeholders participating at both sides, connection recommender systems generally focus on optimizing for source side utility similar to other types of recommender systems \cite{agarwal_chen_2016,liu2007,Adomavicius2005}.
We build on the framework of \cite{Singh_Joachims_2018,Basu_DiCiccio_Logan_El} and formalize the recommendation problem as follows.

Let $s$ denote a source member (initiating a query to the recommender system) and $d$ a destination member (candidate to be shown as a recommendation), and let the implicit query for recommendations $q$ yield an ordered list of $m$ destination members.
The relevance or ranking score for the pair $(s,d)$ is denoted by $u_{s,d}^q$ and reflects an abstract quantity of utility that member $s$ receives from being recommended member $d$.
We assume a ranking policy matrix $P_s^q\in\mathbb{R}^{D_q\times m}$ where $D_q\geq m$ is the total number of eligible destination members for the query and $P_s^q(d,r)$ denotes the probability with which member $d$ is shown to member $s$ in slot $r$ of the recommendation list.
Lastly, $v\in\mathbb{R}^m$ is a fixed vector that models position bias by encoding how much attention destination members pay to recommendations in slots $r=1,\ldots,m$.
This exposure vector is generally chosen to be decreasing and, following previous conventions, we set $v(r)=1/\log(r + 1)$. 
With this setting, we can now define the expected source side utility for member $s$ and query $q$ as
\begin{align*}
    U_q^{s} = \sum_{d=1}^{D_q}\sum_{r=1}^m u_{s,d}^q P_s^q(d,r)v_r = u_s^TP_s^qv.
\end{align*}
We note that the rows and columns of the ranking policy matrix $P_s^q$ sum to at most one and thus the utility-maximizing probabilistic ranking policy can be found by solving the optimization problem
\begin{align}
\label{eq:optprob}
\begin{split}
    \text{arg\,max}_{P_s^q}\ & u_s^TP_s^qv \\
    \text{s.t.} & \sum_{i=1}^{m} P_s^q(d,i) \leq 1\text{ for all } d\in[D_q],\\
    & \sum_{i=1}^{D_q} P_s^q(i,r) = 1\text{ for all } r\in[m],\\
    & 0\leq P_s^q(i,r) \leq 1\text{ for all } i\in[D_q],j\in[m].
\end{split}
\end{align}
Here, the first set on constraints ensures that each destination member is suggested in at most one slot and the second set of constraints requires that each slot of the recommendation is filled with exactly one destination member.
We note that the optimization problem is linear in $D_q\times m$ variables and can thus be solved with standard methods.

If $D_q=m$, the inequalities in the first set of constraints become equalities and the ranking matrix $P_s^q$ is doubly stochastic. 
\cite{Singh_Joachims_2018} use a Birkoff-von-Neumann decomposition to retrieve the deterministic ranking in this case.
However, industry applications generally observe settings with $D_q>>m$, i.e. only a very small subset of all possible members is actually ranked for connection recommendation.
We generate rankings one recommendation slot at a time by iterating over the columns of the ranking policy matrix $P_s^q$ and selecting a row member randomly according to the probabilities denoted in the column. At each iteration step, the rows corresponding to destination members selected for previous recommendation slots are removed and the column values are rescaled to sum to probability one before sampling a new destination member.

\subsection{Adding destination side fairness}

The recommendation procedure described in the previous section optimizes for source side utility without considering the impact on the recommended members which is common practice \cite{agarwal_chen_2016}.
Yet when only a small subset of members can be recommended for any given query, as is the case for many industry-scale connection recommendation systems, exposure in recommendations becomes a scarce resource that can determine who is able to reconnect with old friends or even who receives job opportunities downstream.
In order to understand fairness in recommendations, several metrics have been proposed.

\xhdr{Demographic parity of exposure}
Among the most commonly proposed metrics is demographic parity of exposure \cite[e.g.][]{Zehlike_Castillo_2020,Singh_Joachims_2018,Abdollahpouri_Adomavicius_Burke_Guy_Jannach_Kamishima_Krasnodebski_Pizzato_2019,Singh_Joachims_2019}
which measures the difference in recommendation exposure between groups adjusted for position bias. For a set of disjoint groups of members $G_1,\ldots,G_l$, demographic parity of exposure requires that
\begin{align}
\label{eq:dp}
    \frac{1}{\lvert G_k\rvert}\sum_{d\in G_k}\sum_{r=1}^m P_s^q(d,r)v_r = \frac{1}{\lvert G_k'\rvert}\sum_{d\in G_k'}\sum_{r=1}^m P_s^q(d,r)v_r,
\end{align}
for all $k,k'\in[l]$ which means that groups are displayed in the recommendations at equal rates. 
For two groups $G_0$ and $G_1$, this can be compactly written in vector form as $f^TP_s^qv=0$, where the $d$th entry of $f$ is $f_d=\frac{1(d\in G_0)}{\lvert G_0\rvert} - \frac{1(d\in G_1)}{\lvert G_1\rvert}$.
We note that the constraint is linear and can thus be added to the optimization problem in Equation~\eqref{eq:optprob} without changing the solution approach.

\xhdr{Dynamic parity of utility}
In addition to demographic parity, we consider a dynamic fairness constraint previously referred to as dynamic parity of utility \cite{Basu_DiCiccio_Logan_El}. The constraint requires that different groups receive the same rates of expected utility in each recommendation, i.e. 
\begin{align}
\label{eq:dyn}
\frac{1}{\lvert G_k\rvert}\sum_{d\in G_k}u_{s,d}^q\sum_{r=1}^m P_s^d(d,r)v_r = \frac{1}{\lvert G_k'\rvert}\sum_{d\in G_k'}u_{s,d}^q\sum_{r=1}^m P_s^d(d,r)v_r,
\end{align}
for each $k,k'$ and query $q$. For two groups $G_0$ and $G_1$, we can rewrite this as $\tilde{u}_sP_s^qv=0$ where the $d$th entry of $\tilde{u}_s$ is $(\tilde{u}_s)_d=u_{s,d}^q\left(\frac{1(d\in G_0)}{\lvert G_0\rvert}-\frac{1(d\in G_1)}{\lvert G_1\rvert}\right)$. Note that this is still a linear constraint and can conveniently be added to the optimization problem.
In most applications, the relevance $u_{s,d}^q$ is estimated from data that in some way depends on the current state of the recommender system which dynamically changes the constraint over time, e.g. in connection recommender systems we might use the number of existing connections between members.
Depending on our understanding of $u_{s,d}^q$, the dynamic parity constraint allows for different interpretations.
In our simulation, we will assume that $u_{s,d}^q$ denotes the probability of connection if recommended. In this case, the constraint enforces that destination members in all groups have the same average probability of forming a connection to the source member. If members separate into two groups with shares 2/3 and 1/3 respectively, we would thus expect the recommendation to lead to about twice as many connections to the first group than to the second assuming distributions around the average probability are similar.

\subsection{Scoring model}
\label{sec:scoringmodel}

Connection recommendation requires a notion of relevance of suggestions in order to derive a ranking of members.
In our setting, this is captured by the ranking score $u_{s,d}^q$ for source member $s$, destination member $d$ and query $q$ which reflects the utility member $s$ receives from being recommended member $d$ in query $q$.
In practice, this utility is often modeled by inserting the probability of connection if recommended, some measure of downstream engagement between the two members or a mixture of the two. 
Models for these utility proxies are learned from historic member data and subsequently used to compute ranking scores for new pairs of members.
Since the relevant member data is generally not available to researchers, work in this space often relies on deterministic functions for scoring \cite{Basu_DiCiccio_Logan_El,Singh_Joachims_2018}.
In this paper, we assume that the likelihood with which a member pair $(s,d)$ connects following a recommendation depends on three main characteristics.
First, the larger the current network of $s$, the more likely the member sends invites to recommended members and thus forms connections. This assumption is intuitive since members with large networks tend to be more active in forming connections and in using the platform in general.
Second, the more common connections members $s$ and $d$ have, the more likely they are to connect. This is known as triadic closure in the social networks literature and has been shown to be an important predictor in connection forming \cite{Kossinets_Watts_2006,LibenNowell2007,Krackhardt2007}.
And third, members with a lot of similarities such as similar demographics, interest, education, workplaces, etc. are generally more likely to connect.
This follows the observation that individuals like to be connected to others who are similar to them which is a tendency generally referred to as homophily \cite{McPherson2001,Louch2000,Kossinets_Watts_2006}.

Based on the described components, we assume a model for the connection probability of the pair $(s,d)$ after $d$ has been recommended to $s$ of the form
\begin{align}
    \label{eq:scoringModel}
    \begin{split}
    \log \frac{p}{1-p} = &\ \beta_0 + \beta_1 \text{ networkSize}(s) + \beta_2 \text{ commonConn}(s,d) \\&+ \beta_3 \text{ similarity}(s,d) + \varepsilon,
    \end{split}
\end{align}
where $\varepsilon\sim\mathcal{N}(0,0.1)$ is a random noise term.
We note that the network sizes and numbers of common connections can be easily computed from the adjacency matrix $A_t$ at time $t$, i.e. $\text{networkSize}(s)=1^TA_t(s,\cdot)$ and $\text{commonConn}(s,d) = A_t^2(s,d)$.
For the similarity between members, we assign each member $i$ in our simulation a fixed covariate vector $X_i$ and then set $\text{similarity}(s,d)=-\lvert\lvert X_s - X_d\rvert\rvert_2$.
All features are scaled to lie in $[0,1]$.
To emulate the noise present in data-driven scoring models, we query the connection probability model once to obtain a ranking score for each member pairing and then again to determine if members chosen to be recommended to each other connect which alters the random noise term affecting the score.

\subsection{Simulation procedure}
\label{sec:mainsimprocedure}

We simulate connection recommendation in a fixed size graph of $N=1000$ members with connections evolving over $T=2500$ discrete time steps. Each recommendation consists of a list of $m=20$ ranked individuals which are selected trough the probabilistic ranking framework detailed above.
The frequency with which recommendations are provided to each member are modeled through an exponential waiting times model dependent on the current network size with mean $\lambda=1/(0.001 + 0.02 \times\text{current network size/1000})$.
Separate experiments are conducted for (1) no fairness intervention, (2) demographic parity of exposure intervention, and (3) dynamic parity of utility intervention while resetting random seeds before each intervention type to ensure equal starting conditions.

Members are separated into 65\% majority group (e.g. male members) and 35\% minority group (e.g. female members). We independently sample covariate vectors $X_i\in\mathbb{R}^{30}$ for each member $i$ with $X_i\sim \mathcal{N}(\mu_{G_i},\text{diag}(0.5))$ where $G_i$ denotes the group assigned to member $i$ and $\mu_{G_0},\mu_{G_1}$ are group-dependent means selected randomly from $U([0,1]^{30})$ and fixed throughout all experiments.
The edges of the connection graph are initialized with a stochastic block model with group combination probabilities $(p_0,p_1,p_2)$, i.e. $p_0$ is the probability with which two nodes of group $G_0$ form an initial edge, $p_1$ is the probability with which two nodes of group $G_1$ form an initial edge, and $p_2$ is the probability with which each cross-group pairing forms an initial edge. Group assignments, covariates and the initial graph are fixed for all intervention types in a given simulation run.

For each intervention type, we iterate over the following steps for each $t=1,\ldots,T$.

\begin{itemize}
    \item [(1)] \textbf{Select source members:} We select the members with waiting time zero (source members) and decrease the waiting time of other members by one.
    \item [(2)] \textbf{Score member pairings:} For each source member, we compute the relevance scores to all unconnected members in the graph (destination members) by using the scoring model.
    \item [(3)] \textbf{Solve ranking problem:} A ranking of the destination members is obtained by solving the optimization problem in Equation~\eqref{eq:optprob} subject to fairness constraints if applicable. This is repeated separately for every selected source member.
    \item [(4)] \textbf{Recommendation and addition of connections:} The first $m=20$ members of each recommendation list are suggested to the source member and a connection is formed based on the probabilities obtained by a new call of the scoring model.
    For this, the probabilities are adjusted for position bias and thresholded at 0.5.
    \item[(5)] \textbf{Update parameters:} As a last step, new waiting times are sampled for the source members in this iteration step based on their new network sizes and we repeat the procedure by returning to step (1).
\end{itemize}

\begin{figure*}
    \centering
    \begin{subfigure}[t]{0.23\textwidth}
        \includegraphics[trim = 30 0 10 0, clip=true, scale = 0.09]{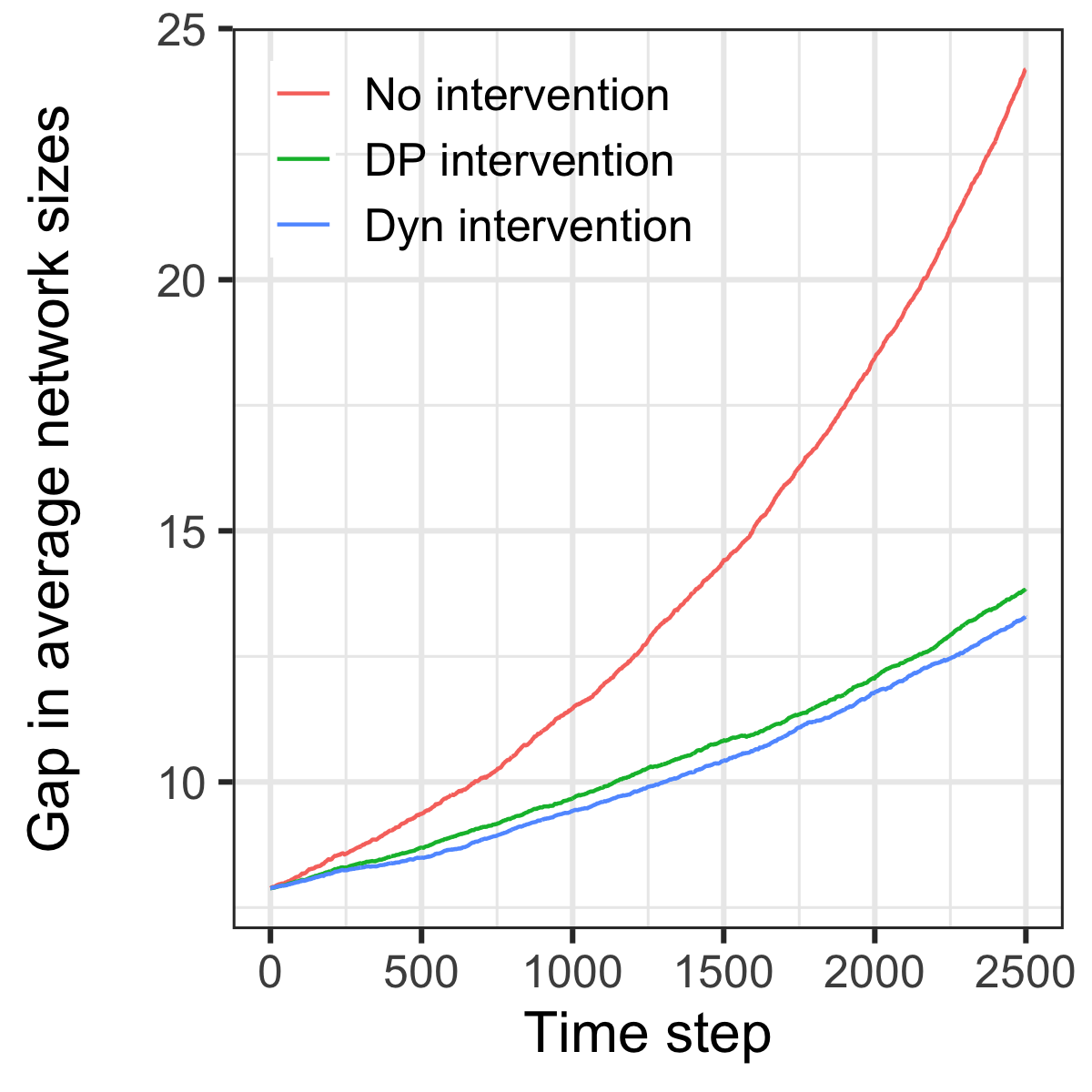}
        \caption{Absolute difference in average network sizes between groups; 0 if average network sizes are the same.}
        \label{fig:gap}
    \end{subfigure}\hfill
    \begin{subfigure}[t]{0.23\textwidth}
        \includegraphics[trim = 30 0 10 0, clip=true,scale = 0.09]{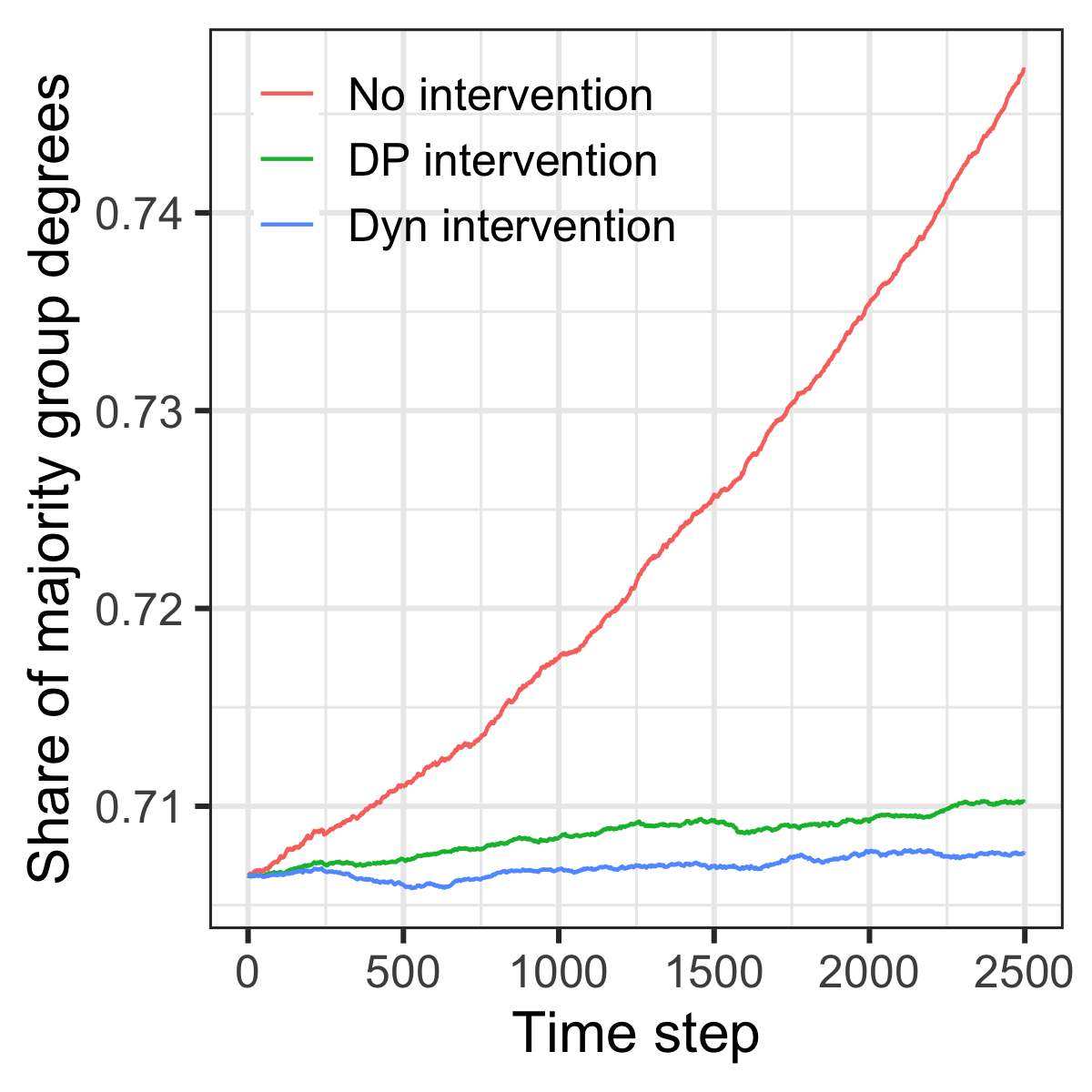}
        \caption{Share of all degrees that belong to majority group; 0.65 if average network sizes are the same.}
        \label{fig:alpha}
    \end{subfigure}\hfill
    \begin{subfigure}[t]{0.23\textwidth}
        \includegraphics[trim = 30 0 10 0, clip=true,scale = 0.09]{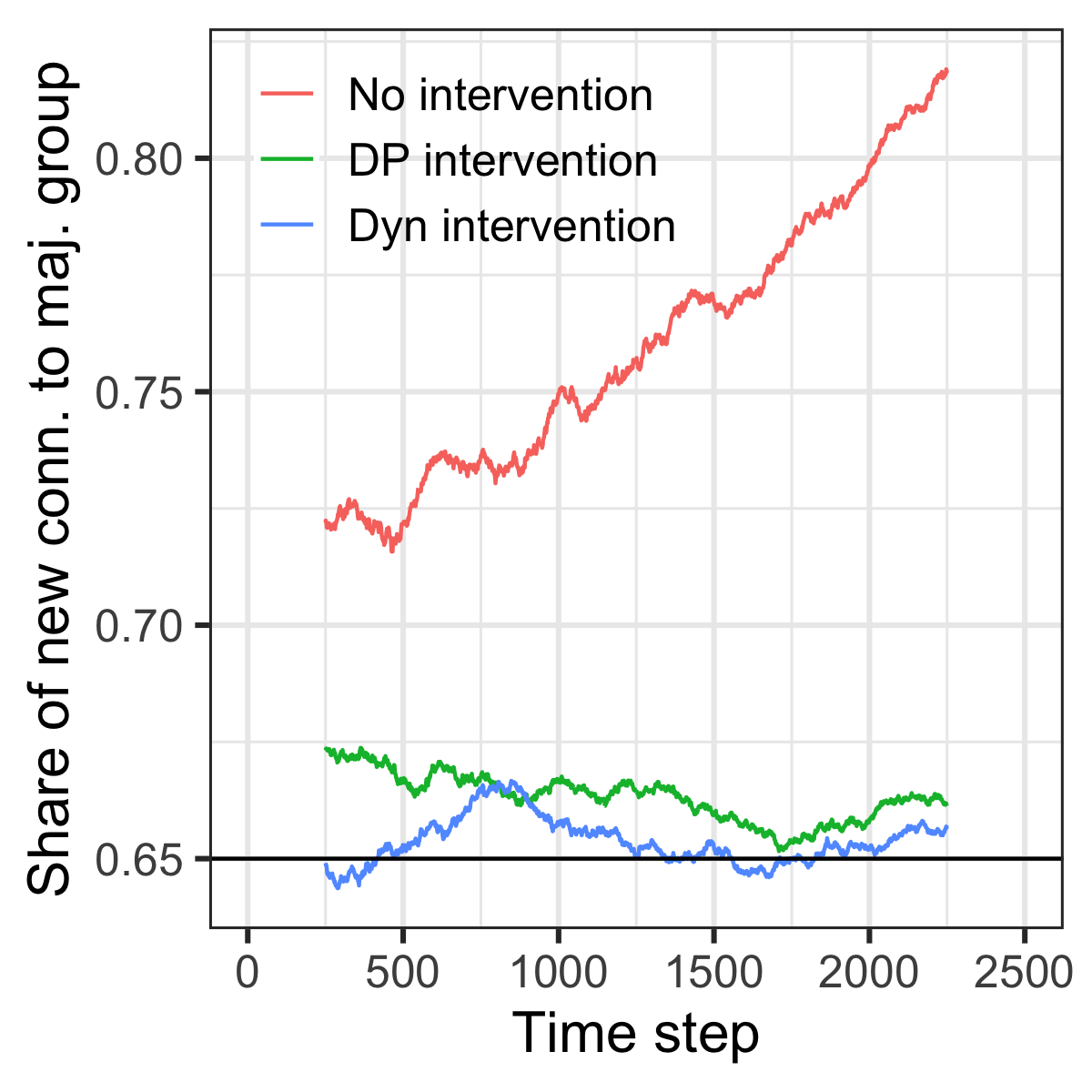}
        \caption{Rolling average (window size 500) of majority group share among destination members of new connections; line at 0.65.}
        \label{fig:share1}
    \end{subfigure}\hfill
    \begin{subfigure}[t]{0.23\textwidth}
        \includegraphics[trim = 30 0 10 0, clip=true,scale = 0.09]{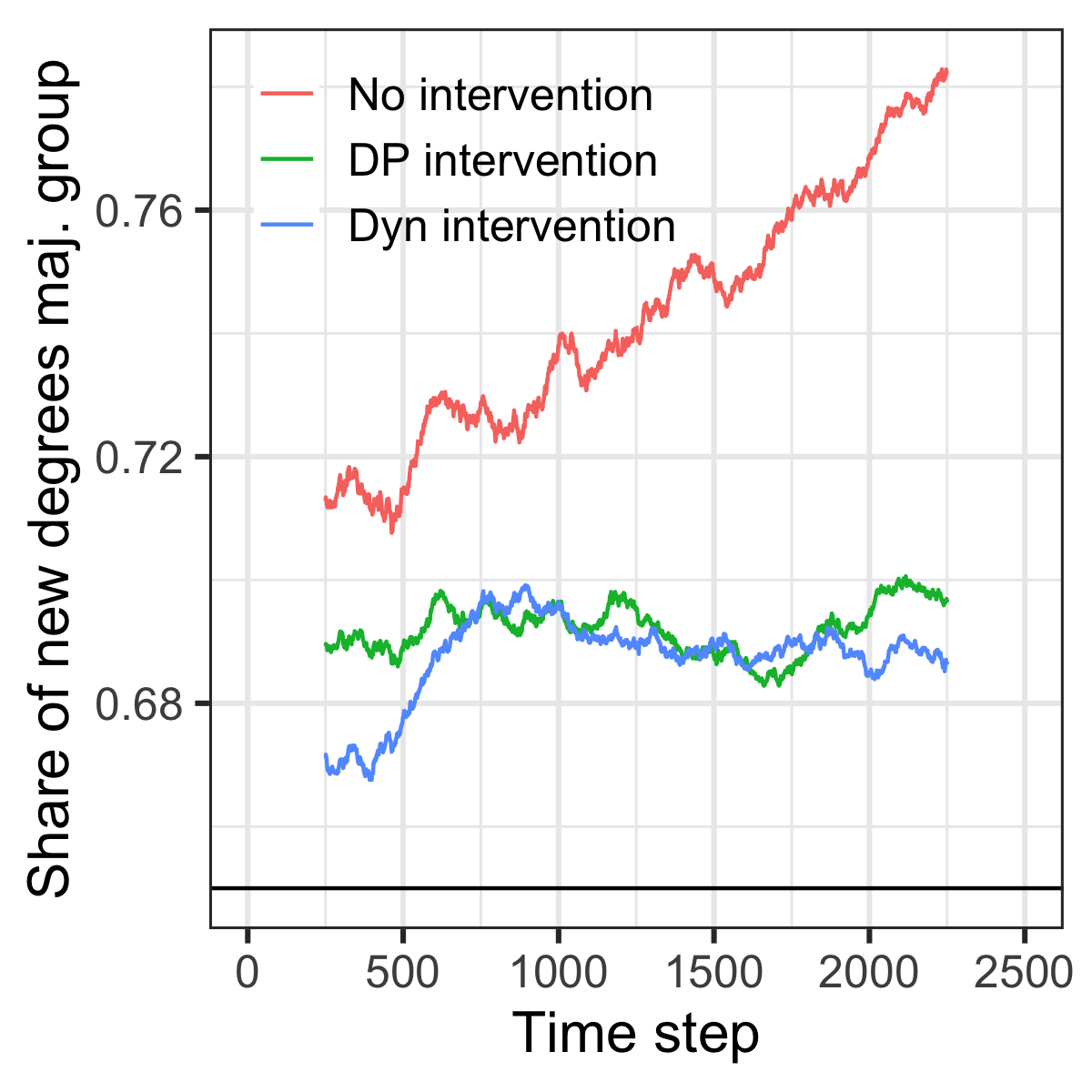}
        \caption{Rolling average (window size 500) of majority group share among new degrees; line at 0.65}
        \label{fig:share2}
    \end{subfigure}
    \caption{Simulation results over 2,500 time steps for no intervention, demographic parity of exposure intervention (DP) and dynamic parity of utility intervention (Dyn). Results are reported as averages over 10 simulation runs. We see that the increase in network size disparities between groups over time is slowed down but not fully mitigated by the fairness interventions.}
    \label{fig:gapalpha}
\end{figure*}

\section{Empirical results}
\label{sec:emp_results}

Experiments are conducted according to the procedure described in Section~\ref{sec:mainsimprocedure} for (1) no fairness intervention, (2) demographic parity of exposure intervention, and (3) dynamic parity of utility intervention. Results are reported aggregated over 10 repetitions of the entire simulation procedure.

\label{sec:mainresults}

\subsection{Graph initialization and scoring model}
We set the stochastic block model parameters for the graph initialization to $(p_0,p_1,p_2)=(0.04,0.032,0.023)$ in order to emulate a realistic setting. This leads to an initial graph in which majority group members have on average 30.16\% more connections than members of the minority group, and members from both groups have more common connections with majority group members than minority group members on average. Average similarity between members of the same group is -3.86 while pairings across groups have an average similarity of -4.35. Although initial feature means differ across groups or group pairings, feature distributions heavily overlap and in-group variations outweigh the differences between groups. Figure~\ref{fig:initial} summarizes the distribution of ranking features at $t=0$.

We set the parameters of the scoring model from Equation~\ref{eq:scoringModel} to $\beta_0=0,\beta_1=50,\beta_2=50$ and $\beta_3=-5$ which has two implications. First, member pairings in the majority group tend to have higher scores than pairings in the minority group because they have larger networks and more common connections, and (2) pairs of members who belong to the same group tend to have higher ranking scores than members from opposite groups because of the different distributions of the number of common connections and the similarity feature. Overall, this leads to decreased ranking scores for minority group members although group membership is not explicitly considered for the computation of scores (Figure~\ref{fig:initial}).

\subsection{Rich-gets-richer in groups}
Figures~\ref{fig:gapalpha}~and~\ref{fig:noReRanker} depict the results of the connection recommendation experiment if no fairness intervention is applied. 
Over time, the initial gap in average network sizes between groups increases as majority members are able to grow their networks faster than members of the minority group ( Figure~\ref{fig:gap}). In addition, the majority group share among new connections is increasing over time ( Figures~\ref{fig:share1} and \ref{fig:share2}) which leads to a superlinear growth of the network size gap and suggests a positive feedback loop which amplifies the advantage of the majority group over time. 
Starting out with on average 30.18\% larger networks, members in the majority group have on average 59.88\% more connections after $t=2500$ time steps.
Figure~\ref{fig:noReRanker} shows that the distribution of network sizes at $t=2500$ follows a power law distribution with a particularly long tail for majority group members and lower mode for the minority group suggesting that most majority members have larger networks than most minority members. In addition, the figure depicts the the relation of initial network sizes and network sizes at $t=2500$ on a log-log scale. The additional curves display the network size of individuals in a counterfactual scenario in which the growth of networks within a group is uniformly distributed among all member of the group, i.e. the curves correspond to $f(\text{network size at } t=0)=(\text{average increase in } G_i) + (\text{network size at } t=0)$ on the log-log scale. The result suggests that members who grow their networks more than the average within their group in the given time frame tend to be the members who had larger networks to begin with.

In summary, our key findings from simulation with no fairness intervention are: (1) Unconstrained connection recommendation increases the initial disparity in average network sizes befitting the already advantaged majority population. (2) Network sizes tend to a power law distribution with a lower mode for the minority population. (3) The members whose network sizes are in the tail of the power law distribution tend to be the members who had large networks as compared to the rest of their groups to begin with. Overall, these observations confirm a `rich-get-richer' or Matthew effect where majority members appear to benefit more from the phenomenon than minority members.

\subsection{Demographic parity of exposure intervention}
We conduct the same experiments with demographic parity fairness intervention and present the results in Figures~\ref{fig:gapalpha} and \ref{fig:dpdyn}. While majority group members are overexposed with an exposure share of 75.5\% without intervention, the demographic parity intervention leads to a majority group exposure share of 66.3\% averaged over all time steps and simulation runs which is close to the 65\% population share of the majority group. While this suggests that the intervention fulfills its purpose in aggregate, Figures~\ref{fig:gap} and \ref{fig:alpha} show that network sizes are not converging as intended. Although the intervention leads to less outcome disparity than in the unconstrained setting, both the gap in average network sizes and the share of degrees that belong to the majority group increase over the period of the experiment. This is because (1) majority group members in our experiment seek out connection recommendations more frequently, and (2) majority members have higher ranking scores and higher likelihoods to connect. Both of these points lead to more than 65\% of new connections being formed to and from majority group members (Figures~\ref{fig:share1} and \ref{fig:share2}) which exacerbates the differences in network sizes instead of mitigating them. On a high level, the demographic parity of exposure intervention ensures that members of different groups are displayed in recommendations at the same rates which does not lead to equal connection rates when the underlying relevance distributions vary.
At $t=2500$, network sizes in both groups appear to follow a power law distribution with lower mode in the minority group (Figure~\ref{fig:dpdyn}) which means that the majority of members is still disadvantaged. As compared to no intervention, the average network size of minority group members at $t=2500$ increases by 2.11 with median increase of 1.

Overall, the results show that enforcing demographic parity of exposure in recommendation lists is not sufficient in order to achieve parity of average network sizes between groups. 
Although seemingly fair in aggregate, the gap in average network sizes is still increasing over time. This growth is happening at a much slower rate than without fairness intervention suggesting that (part of) the bias amplification feedback loop in the dynamic system is mitigated. We theoretically examine the workings of the intervention in a stylized model in Section~\ref{sec:theory}.

\subsection{Dynamic parity of utility intervention}
Results for the dynamic parity of utility case are depicted in Figure~\ref{fig:gapalpha} and \ref{fig:dpdyn}.
The distribution of network sizes after $t=2500$ time steps of the experiment closely resembles the distribution for the demographic parity of exposure case with generally lower network sizes in the minority group and a median increase of 1 connection as compared to the results of the unconstrained connection recommendation. On average, minority group members gain 2.6 more connections than without intervention. 

We observe that the absolute gap in average network sizes and the share of majority group degrees are increasing. However, the increase is happening at a slower rate than in the demographic parity of exposure setting. Figure~\ref{fig:share1} shows that the majority group share among the destination members of new connections hovers around the desired 0.65 mark (on average 0.654) which suggests that the intervention successfully ensures that members of both groups have about the same average probability to gain connections through being displayed in recommendations to other members. However, the majority group share among all new degrees exceeds the desired share and averages to 0.687 over all simulation runs and time steps (Figure~\ref{fig:share2}) which leads to an increasing gap in average network sizes. This is because (1) majority group members seek out recommendations more frequently based on their larger network sizes, and (2) source members who belong to the majority group are able to connect to more of the recommended members since they generally have larger scores and the parity of utility within a single recommendation list does not imply parity of total utilities between recommendation lists for different source members. 
While the dynamic parity of utility intervention solves some of the problems we observed with the demographic parity of exposure intervention, it suffers from the same limitations regarding the biases introduced by the source side of the connection recommendation system. Our theoretical derivations in Section~\ref{sec:theory} suggest that the dynamic parity of utility intervention can lead to stably fair average network sizes in setting with no source side bias.

\section{Theoretical characterization}\label{sec:theory}

\subsection{Urn models and mixed preferential attachment}
\label{sec:mpa}

\xhdr{Urn models for dynamic systems of unfairness}
The results of the simulation study in Section~\ref{sec:emp_results} demonstrate how fairness intervention in connection recommender systems can lead to unanticipated long-term effects.
In order to understand why these effects occur and how to reach a fair balance of network sizes, we seek out a theoretical analysis of the impact of intervening on the connection recommendation dynamics.
While our simulation setup resembles the workings and data setting of real-world connection recommender systems, it is quite complex and a full theoretical characterization of the behavior of the system requires major simplifications to the model structure. 
Urn models present a class of models whose behavior is more tractable to analyze in theory but that still proves flexible enough to lend itself to various applications \cite{Pemantle_2007}.
They have been previously used in the algorithmic fairness literature to model feedback loops in predictive policing \cite{Ensign_Friedler_Neville_Scheidegger_Venkatasubramanian_2018}.

For the connection recommendation purpose, we employ a type of dynamic growth urn which is also known as preferential attachment model \cite{Barabasi1999}.
Preferential attachment models are generative random network models which rely on a local growth rule that renders vertices that already have a large number of connections likely to accumulate more connections over time similar to the setting in our simulation study. %
In an effort to extend the preferential attachment setting to social networks with members of different groups, researchers have proposed a mixed preferential attachment model that allows for a majority-minority partition and consideration of homophily \cite{Avin2015,Avin_Daltrophe_Keller_Lotker_Mathieu_Peleg_Pignolet_2020}.
We draw on this type of model to theoretically characterize the workings of fairness intervention in the connection recommendation setting and formally define the mixed preferential attachment model specification we use in the following.
Note that, while the models considered in the empirical and theoretical parts of this work are not the same, they are similar enough to warrant the expectation that some of the qualitative observations from the analysis of the mixed preferential attachment model can be translated to insights into the behavior of the realistic simulation study on a group-aggregate level. Appendix~\ref{sec:app_comparison_models} summarizes the similarities and differences between the two models.

\xhdr{Mixed preferential attachment (MPA) model}
Let $\mathcal{G}_t(r,d_0,\pi)$ be a bi-populated evolving random graph with nodes in groups $G_0$ and $G_1$.
Here, $d_0\in\mathbb{N}$ is the sum of all degrees at $t=0$, $r\in (0,0.5]$ is the arrival rate of the $G_1$ vertices, and $\pi\in\mathbb{R}^{2\times 2}$ is the mixing matrix.
For simplicity of proofs, we assume the fraction of initial vertices that belong to group $G_1$ is $r$.
We denote degree of a vertex $v$ at time $t$ as $d_t(v)$, the sum of all degrees as $d_t$, and the sum of all degrees within groups $G_0$ and $G_1$ as $d_t(G_0)$ and $d_t(G_1)$ respectively.
Note that each connection in the graph translates to two degrees, one for the node at either side of the connection.
The generative process of the graph works as follows.
In each iteration $t$, a new node, which belongs to group $G_1$ with probability $r$ and to $G_0$ otherwise, is added to the network.
The new node can be interpreted as the source member who seeks out a connection recommendation and is subsequently connected to exactly one existing node in a two-stage recursive procedure.
First, we sample a tentative neighbor at random with probabilities proportional to the degrees of the existing nodes at time $t$, i.e. $P(\text{select node }v)=d_t(v)/d_t$. This member corresponds to the recommended destination member. 
Second, we denote the groups of the new and selected nodes and sample whether the connection is successful based on the mixing matrix
$$
    \pi = \begin{bmatrix}p_0 & 1-p_0\\ 1-p_1 & p_1\end{bmatrix}.
$$
If both nodes belong to group $G_0$, the connection is successful with probability $p_0$. If the first node belongs to group $G_0$ and the second node is from group $G_1$, the connection is successful with probability $1-p_0$, etc.
This step can be interpreted as the members' reaction to the recommendation where the row vectors $\pi_{i}=(p_i, 1-p_i)$ represent the homophily preferences of the groups, i.e. how much members prefer connections within their own group over connections to members of the other group. 
When $p_i\in (0.5,1)$, members are assumed to be positively biased towards connections within their own group (homophily), and for $p_i\in (0,0.5)$, members prefer connections to the other group (heterophily).
If the connection fails, a new recommendation is made by sampling a new tentative neighbor and repeating the procedure until the new node connects to exactly one existing node.

\subsection{Rich-gets-richer in groups}

It is well known that the degree distribution of nodes in preferential attachment models tends to a power law distribution leading to a `rich-get-richer' phenomenon \cite[e.g.][]{linyuan2006complex}. The authors of \cite{Avin_Daltrophe_Keller_Lotker_Mathieu_Peleg_Pignolet_2020} extend this result to social network settings where each node belongs to one of two groups. 
Let $\alpha_t = d_t(G_1)/d_t$ be the rate of minority degrees in a mixed preferential attachment network at time $t$. Then, the paper shows that there is a limit $\alpha$ independent of the initial graph such that $\lim_{t\to\infty}\mathbb{E}[\alpha_t]=\alpha$.
Letting $m_{k,t}(G_i)$ denote the number of vertices in group $i$ with degree $k$ at time $t$ and $M_k(G_i)=\lim_{t\to\infty}\mathbb{E}[m_{k,t}(G_i)]/t$, the paper follows that the numbers of degrees in groups tend to power law distributions $M_k(G_i)\propto k^{-\beta(G_i)}$ with different exponents 
\begin{align*}
    \beta(G_0) = 1 + \frac{1}{c_0} \text{\ \ \ \ and\ \ \ \ } \beta(G_1) = 1 + \frac{1}{c_1},
\end{align*}
where
\begin{align*}
    c_0 &= \frac{1}{2}\left(\frac{(1-r)p_0}{p_0+\alpha-2p_0\alpha} + \frac{r(1-p_1)}{1-p_1-\alpha+2p_1\alpha}\right), \text{ and }\\
    c_1 &= \frac{1}{2}\left(\frac{(1-r)(1-p_0)}{p_0+\alpha-2 p_0\alpha} + \frac{rp_1}{1-p_1-\alpha+2p_1\alpha}\right).\\
\end{align*}
Both exponents are functions of the rate of the minority group $r$, the limit $\alpha$ and the mixing matrix $\pi$.
In the case of perfect homophily $p_0=p_1=1$ or no homophily bias $p_0=p_1=0.5$, the exponents $\beta(G_i)$ are the same for both groups and no group is disadvantaged in the long term.
In the more realistic setting $p_0,p_1\in (0.5,1)$, members are more likely to connect to members in the same group but have a non-zero probability of connecting to the members of the other group if recommended. %
If $p_0\geq p_1$ in this setting, one can show that $\beta(G_0)\geq\beta(G_1)$ and thus the networks of majority group members outgrow the networks of minority group members in the long run.
If $p_1>p_0>0.5$, the picture is more complex, and networks of the minority group can outgrow the networks by the majority group in cases in which the difference between $p_1$ and $p_0$ is large or the rate of the minority group $r$ is close to 0.5.

In the setting corresponding to our simulation study, we have $r=0.35$ and $p_0=p_1>0.5$ since the similarity feature in scoring model uses the same parameter for both groups.
The MPA model suggests that in this case the network sizes in groups tend to a power law distribution with larger networks in the majority group which aligns with our empirical observations. 
We set $r=0.35$ in the MPA model and compute the analytical limits of the expected share of minority group degrees for different combinations of $p_0$ and $p_1$.
The results in Figure~\ref{fig:alphas} (left plot) confirm that (1) most combinations of $p_0$ and $p_1$ lead to divergent network sizes in the long run, and (2) in the homophily setting with $p_0,p_1>0.5$ the majority group is more likely to benefit while in the heterophily setting with $p_0,p_1<0.5$ the minority group is more likely to be advantaged.
For our setting with $p_0=p_1>0.5$, the figure shows that the minority group share of all degrees in the network remains smaller than the desired 35\% in the long run which means minority group members are left smaller networks in general.

\subsection{Enforcing parity in recommendations}
\label{sec:theoryDP}

\begin{figure*}
    \centering
    \includegraphics[trim = 30 0 10 0, clip=true, scale = 0.09]{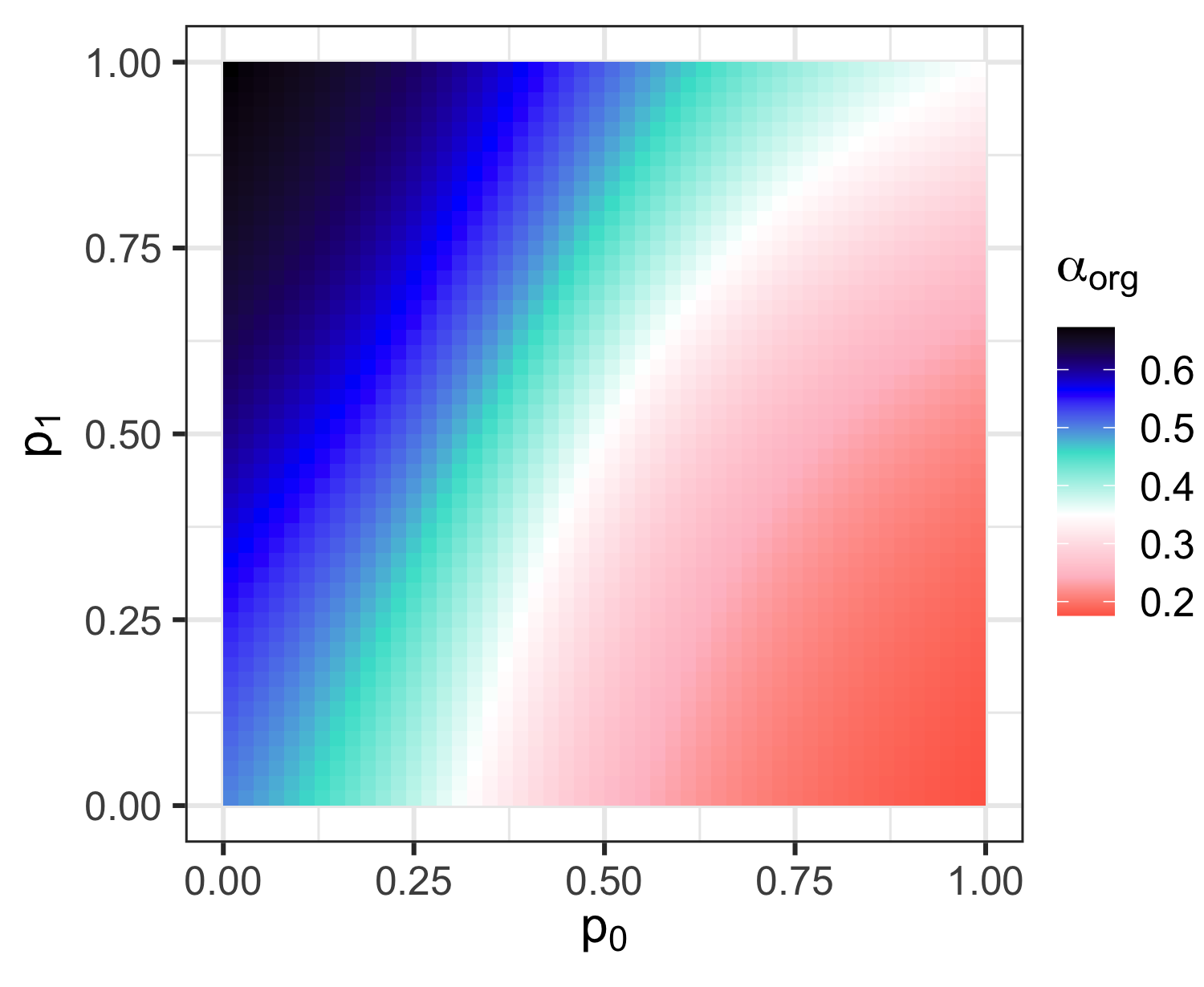}
    \includegraphics[trim = 30 0 10 0, clip=true,scale = 0.09]{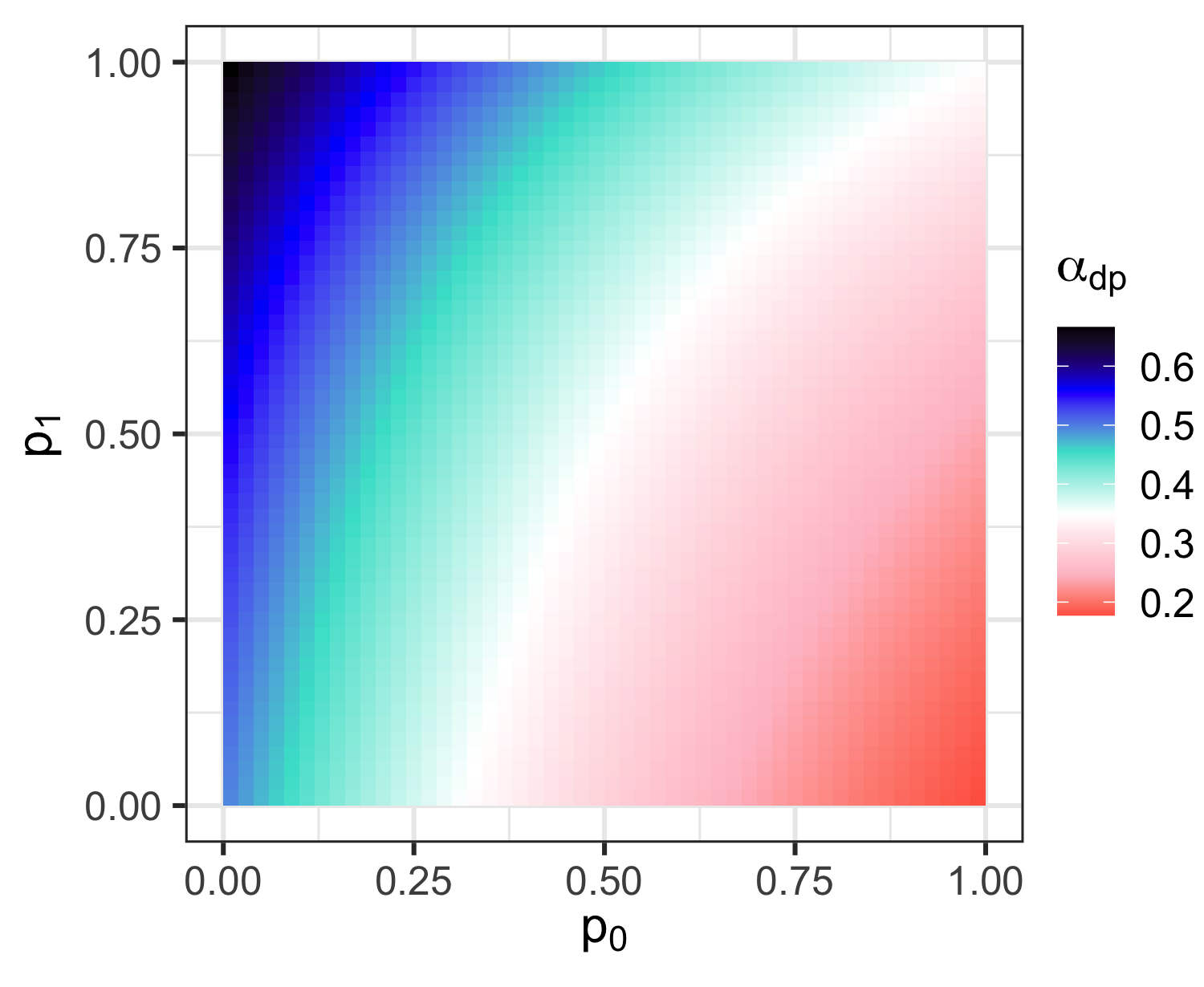}
    \includegraphics[trim = 30 0 10 0, clip=true,scale = 0.09]{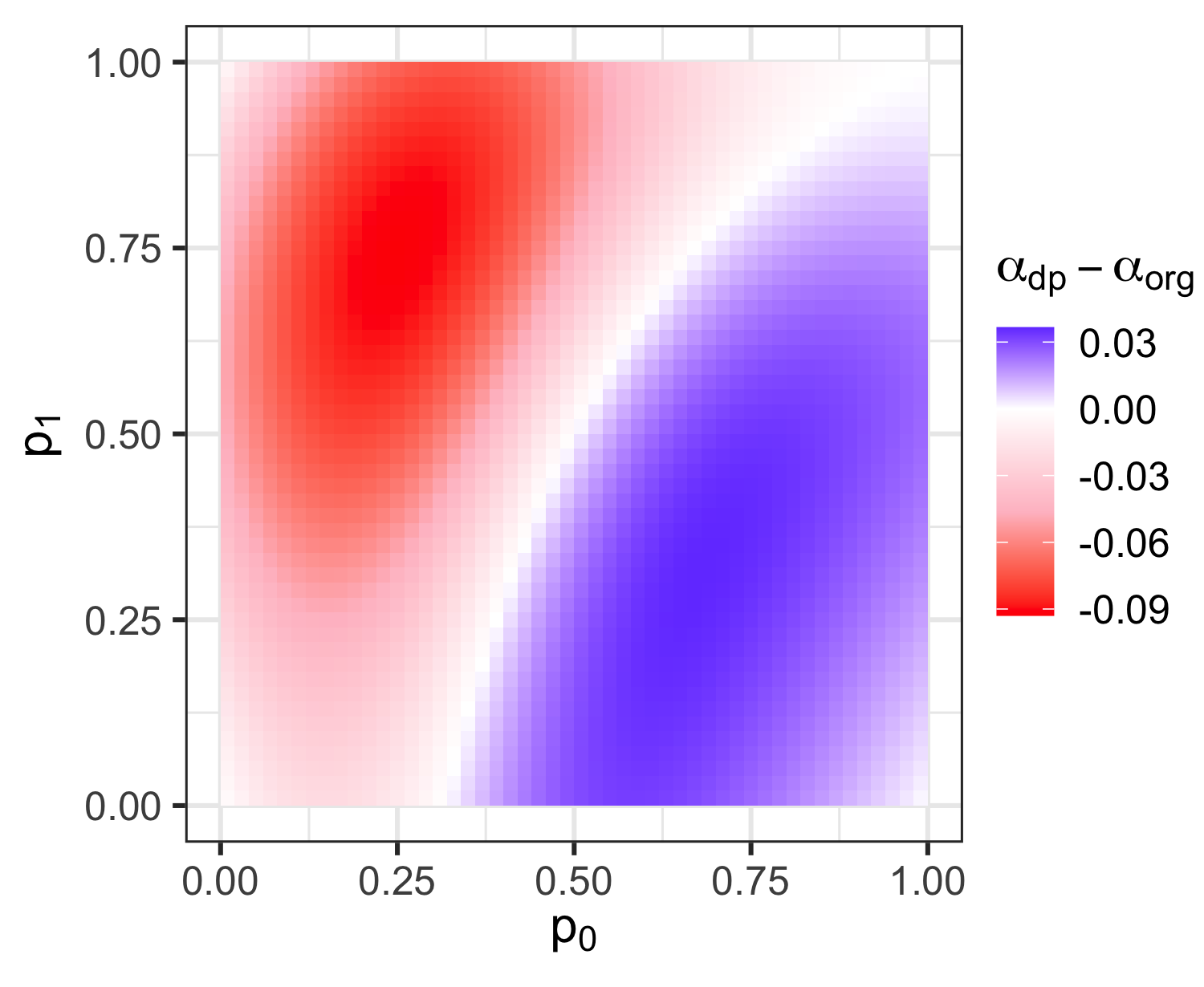}
    \caption{Analytical limits $\alpha=\lim_{t\to\infty}\mathbb{E}[\alpha_t]$ for $r=0.35$ and $p_0,p_1\in(0,1)$. In the left and middle plots, $\alpha=r=0.35$ (white) indicates that the average network size in $G_0$ and $G_1$ is the same in the long term, $\alpha<0.35$ (pink\& red) indicates that members of group $G_0$ have larger average networks and $\alpha>0.35$ (blue\& black) that members of $G_1$ have larger networks. The left plot depicts the limit in the original mixed preferential attachment model, the middle plot the limit with demographic parity intervention, and the right plot the difference between the two. We see that network sizes in groups diverge with and without intervention for most combinations of $p_0$ and $p_1$. While the demographic parity intervention does not correct the entire outcome disparity, it does lead the limit $\alpha$ closer to $r$ as depicted in the right plot. The homophily setting with $p_0,p_1\in(0.5,1]$ corresponds to the upper right quadrants of the plots.}
    \label{fig:alphas}
\end{figure*}

We saw in our empirical results that enforcing demographic parity of exposure slows down the increase of network size disparities but does not lead to similar distributions of network sizes between groups over time.
To understand why this is the case, we analyze the effect of a parity of exposure intervention in the MPA model. 
The demographic parity of exposure condition in Equation~\eqref{eq:dp} requires that the average expected exposure in rankings is the same for members of both groups. Since the MPA model recommendations only have one slot for each query, there is no consideration of position bias in this setting and $v_1=1$. 
The fairness condition becomes
$$
    \frac{1}{\lvert G_0\rvert}\sum_{d\in G_0}P(\text{recommend }d) = \frac{1}{\lvert G_1\rvert}\sum_{d\in G_1}P(\text{recommend }d).
$$
Recall that in the original setting of the MPA model, the incoming source member at time $t$ belongs to minority group $G_1$ with probability $r$, and the recommended destination member is selected from the existing graph with probability determined by their network size $P(\text{select node }v)=d_t(v)/d_t$.
In order to fulfill the fairness constraint, we alter the mechanism with which the destination member is sampled and first, sample the group from which to choose a member $G_i$ with $i\sim\text{Bin}(r)$, and second, sample a destination member from that group according to their network sizes with $P(\text{select node }v)=d_t(v)/d_t(G_i)$ for all $v\in G_i$.
As before, the two members are connected with probabilities determined by the mixing matrix and the procedure is repeated until the source member forms exactly one connection. 
We characterize the limiting rate of minority group degrees in this setting in the following theorem. A proof can be found in Appendix~\ref{sec:proof_dp}.

\begin{theorem}
\label{thm:dp}
Assume a mixed preferential attachment model with demographic parity intervention as described. Then for $t\to\infty$, the share of degrees of the minority group
$\alpha_t=d_t(G_1)/d_t$ tends to a fixed value $\alpha=\alpha(\pi,r)$ which is independent of the initial graph. Specifically, 
$$
\alpha := \lim_{t\to\infty}\mathbb{E}[\alpha_t] =
\frac{1}{2}\left(r\frac{rp_1}{1-r-p_1+2rp_1} - (1-r)\frac{(1-r)p_0}{r+p_0-2rp_0} + 1\right).
$$
\end{theorem}

If the demographic parity intervention would lead to the same average network sizes in groups in the long-term, we would have $\lim_{t\to\infty}\mathbb{E}[\alpha_t]=r$ which is exactly the fraction of $G_1$ nodes in the graph. However, the limit in Theorem~\ref{thm:dp} does not equal $r$ in general.
A sufficient condition for $\alpha=r$ is the trivial case with $r=0.5$ and $p_1=p_0$ in which no bias from the in-group connection probability and the different group sizes is introduced in the first place.
However, one can show that in the homophily regime of $p_0,p_1>0.5$ and with $r\in(0,0.5)$, the case $p_0>=p_1$ leads to $\alpha<r$ and the case $p_1>p_0$ can lead to $\alpha>r$ if $p_1-p_0$ is large or $r$ is close to 0.5 similar to the results without intervention. 
Figure~\ref{fig:alphas} displays the limit $\alpha$ after demographic parity intervention for different combinations of $p_0$ and $p_1$ at $r=0.35$ (middle plot). We see that for only very few combinations of $p_0$ and $p_1$ the limit aligns with the desired share $r=0.35$. In fact, the left plot shows that those combinations align exactly with the values that lead to fair outcomes even without intervention.
Although the demographic parity intervention cannot fully neutralize the bias in network sizes, the figure also shows that it moves the solution $\alpha$ closer to the desired solution $r$ which diminishes the limiting gap in average network sizes. 
On a high level, the demographic parity intervention can only ensure that members from different groups are recommended at the same rates which is not sufficient to address the asymmetries introduced by the different network sizes and probabilities to connect after recommendation within and in-between groups. The intervention essentially interrupts the group-wise feedback loop stemming from faster growing networks in one of the groups while leaving other sources of bias untouched. 

These theoretical results align with the empirical results from the simulation study which show that demographic parity intervention does not lead to equality in average network sizes over time, and minority group members are still largely disadvantaged. Recall that the regime of the simulation study translates to $r=0.35$ and $p_0=p_1>0.5$ to see this.
A key difference between the main simulation and the MPA model are the ground truth probabilities of connection after recommendation. In the MPA model, these probabilities are fixed constants for each of the four combinations of source and destination groups while the simulation model relies on probabilities which are positively affected by larger networks of the source and destination members.
In both models, the demographic parity intervention corrects the group-wise feedback loop from larger networks to a higher chance at being selected for recommendation. However, the feedback loop in the main simulation does not only affect the chance of being recommended but also the chance of connection after recommendation which is unaffected by the intervention.
This could explain why the disparity growth in Figure~\ref{fig:gapalpha} appears to remain superlinear even with fairness intervention.

\subsection{Stable equality of average network sizes}
\label{sec:dyn}

Both the main simulation study and the analysis of the MPA model show that enforcing demographic parity of exposure in connection recommendation lists is generally not sufficient to reach an equilibrium of equal average network sizes between groups. In the following, we explore what type of intervention is needed to reach this parity state. 
We recall that in order for the average network sizes to be equal between the two groups in the long-term, the limiting share of minority group degrees $\alpha=\lim_{t\to\infty}\mathbb{E}[\alpha_t]$ needs to be $r$.
An easy way to achieve this would be to effectively set the probability of cross-group connections to zero by introducing an additional rejection sampling step that skips a tentative neighbor and resamples whenever the group differs from the group of the source member. 
Of course in practice, a solution which only allows in-group connections is undesirable.
Yet the same rejection sampling idea can be used to derive a more desirable intervention mechanism if we allow for the adjustments to be dynamic. 

Consider the following alteration of the mixed preferential attachment iteration step described in Section~\ref{sec:mpa}. Like before, a new member enters the graph and with probability $r$, the new member belongs to group $G_1$. We then select a tentative destination member from the graph with probabilities determined by the network sizes of the existing members. Now, we insert a new rejection sampling step in which the tentative destination member is retained with probability $q_{ij}$ where $G_i$ is the group of the source member and $G_j$ the group of the destination member. If the proposal is rejected, a new tentative destination member is selected as before until a member can be retained successfully. Only then is the destination member recommended to the source, and an edge is inserted with probability determined by the mixing matrix $\pi$. If the connection is unsuccessful, the whole procedure is repeated until the sampled source member connects to exactly one existing member. The following theorem characterizes how the probabilities $q_{ij}$ have to be selected in order to obtain a stable equality of average network sizes between groups. A proof for the theorem is given in Appendix~\ref{sec:proof_dyn}.

\begin{theorem}
\label{thm:dynamicintervention}
Assume a mixed preferential attachment model with additional rejection sampling step as described above. Let $q_{ij}$ denote the probability with which we retain a tentative destination member of group $G_j$ as possible connection recommendation to a source member of group $G_i$. For each iteration step $t$, we set
\begin{align*}
\begin{alignedat}{2}
q_{00}&=\frac{(1-r)(\alpha_t(p_0-2)+2)}{p_0(\alpha_t-r)}, \hspace{10ex} &q_{01}=1-q_{00},\\
q_{11}&=\frac{(1-\alpha_t)(1-p_1)r}{\alpha_t(p_1-r)-p_1r+r}, \hspace{10ex} &q_{10}=1-q_{11},
\end{alignedat}
\end{align*}
and map values outside of $[0,1]$ to $0$ and $1$ respectively.
The $q_{ij}$ are selected such that the sampled source member connects to a member of group $G_1$ with probability $r$ and to a member of group $G_0$ with probability $1-r$ in every iteration step $t$, and thus it holds that $\lim_{t\to\infty}\mathbb{E}[\alpha_t]=r$.
\end{theorem}

An important insight from this result is that, in order to ensure stable equal average network sizes between groups in a realistic fashion, we require a dynamic type of fairness intervention that changes with the state of the system. In our case, the probabilities with which a tentative recommendation needs to be rejected depends on the share of minority group degrees in the network $\alpha_t$ and changes as this share evolves.
The dynamic intervention is fundamentally different from the demographic parity intervention considered previously.
While the demographic parity constraint ensures that members from both groups have the same average probability of being recommended to a source member, the dynamic parity procedure arranges that the probability of connecting to a destination member of the minority group $G_1$ is always $r$ all things considered. This balances out the potential biases introduced on the destination side of the recommendation.

In our experimental setup, an intervention with the described effect is given by the dynamic parity of utility case. Like the rejection sampling idea in the MPA model, dynamic parity of utility ensures that the probability of connecting to members of both groups is proportional to the population group shares in any given recommendation list. Different to the MPA model, additional bias is introduced through the source side in our simulation framework which more closely resembles real-world settings. The theory presented here suggests that dynamic parity of utility intervention can lead to stably fair average network sizes in setting with no source side bias and similar score distributions across groups.

\section{Discussion}
\label{sec:discussion}

\xhdr{Findings}
We analyze long-term dynamics of fairness intervention in connection recommender systems by (1) studying a simulation-based recommender system patterned after
the systems employed by web-scale social networks, and (2) theoretically analyzing how certain interventions on fairness impact the bias amplification dynamics in stylized connection recommendation using mixed preferential attachment models.

Our empirical and theoretical findings suggest that unconstrained connection recommendation leads to amplification of initial differences in average network sizes between groups, and a group-wise rich-get-richer effect benefiting the majority population and especially majority group members who had relatively large networks to begin with which is in line with previous research \cite{Chen_Dong_Wang_Feng_Wang_He_2020,Yao_Halpern_Thain_Wang_Lee_Prost_Chi_Chen_Beutel_2021}.
We find that intervening by enforcing demographic parity of exposure in recommendation lists as commonly suggested in the literature \cite[e.g.][]{Zehlike_Castillo_2020,Singh_Joachims_2018,Abdollahpouri_Adomavicius_Burke_Guy_Jannach_Kamishima_Krasnodebski_Pizzato_2019,Singh_Joachims_2019}
leads to less bias amplification but is not sufficient in order to mitigate an increase in the disparities in network sizes over time. Although seemingly `fair' in aggregate, most minority group members remain disadvantaged in the long run.
Moving to dynamic parity of utility intervention alleviates some of the problems posed by the demographic parity of exposure case but still results in increasing disparities over time. This is %
because fairness is only evoked in individual recommendation lists and does not affect bias introduced through the source side of the recommendations.%

Most commonly, the efficacy of fairness intervention in recommendation is measured by a single fixed fairness criterion that is evaluated in a one-shot or time-aggregate manner \cite{Singh_Joachims_2018,Zehlike_Bonchi_Castillo_Hajian_Megahed_Baeza-Yates_2017,Zehlike_Castillo_2020}. Our work demonstrates how this can lead to deployment of fairness enhancing algorithms with unforeseen consequences in the long run by hiding variations in fairness and other metrics over time. Ultimately, connection recommendation operates on a dynamical system which needs to be taken into account explicitly in order to ensure equitable outcomes in the long run.

\xhdr{Sensitivity to source side bias}
Theoretical analysis of our urn-based model suggests that dynamic parity of utility intervention can mitigate disparities in network sizes if and only if no bias is introduced through the source side of the recommendation process.
Yet, source side bias is to be expected in real-world settings and our simulation study opts to incorporate such bias in several ways: (1) We assume that users with larger networks are served connection recommendations more frequently. %
(2) We assume that users with larger networks are more likely to form connections based on recommendations. %
(3) We assume initial differences in the distributions of users' similarity and their number of common connections.%

The last point is build around the homophily and triadic closure ideas discussed in Section~\ref{sec:scoringmodel} and leads to more source utility and thus more new connections per recommendation for the majority group. 
The first two mechanisms follow a more intuitive rationale. We generally assume a positive correlation between network size and platform activity levels of users which naturally leads users with larger networks to come across more recommendations for connections. Given that those users have large networks, we assume that they are somewhat proactive in forming ties which leads them, on average, to form more connections per recommendation than users with smaller networks. 
Both of these assumptions have to be carefully checked in practice and might not hold true in all settings, e.g. one could imagine a scenario in which a user with very large network stops to proactively seek out connections based on recommendations instead relying on connection invitations from others. However, as long as some sort of source side bias between groups is introduced, our findings suggest that the studied types of fairness interventions are not sufficient to prevent outcome disparities and targeted source side fairness interventions are needed.

\xhdr{Moving towards real-world impact} Most research in fair recommendation abstracts away from specific application settings %
which has been criticized as ineffective and in some cases even harmful \cite{patro2022fair,Selbst2019}.
Instead, our work assumes a concrete connection recommendation setting patterned after the systems employed by real-world social networks which allows us to draw concrete conclusions for application. 
Nevertheless, assessing the exact impact and possible side effects of fairness intervention after deployment in real-world systems remains difficult because of additional complexity and noise \cite{Holstein_Wortman_2019}.
One component that often remains unaddressed is the role of user feedback. In many cases, destination side recommendation utility and fairness are measured by using ranking exposure as a proxy variable which ignores potential variations in user response \cite{patro2022fair}. 
In settings where users directly discriminate against one group, increasing the group's exposure in recommendations as part of a fairness enhancement effort could even lead to adverse outcomes for the group as demonstrated in other settings \cite{Liu_Dean_Rolf_Simchowitz_Hardt_2018,lifeed}.
Understanding and modeling the role of human biases in connection recommendation is an integral extension of the work presented here.
Further complications are introduced by the noise and uncertainty in real-life recommendation settings. For example, relevance scores can usually only be estimated from data plagued by selection bias which introduces noise and additional biases into the system \cite{Emelianov2020,patro2022fair}. In order to obtain balanced data, we would have to employ a uniformly at random recommendation policy which has been attempted by researchers in the past \cite{Jiang2019,Liu2020} but inevitably hurts the experience of members \cite{Chen_Dong_Wang_Feng_Wang_He_2020}.
In addition, most fairness measures assume access to individual level demographic information which can be hard to obtain in practice for legal reasons and concerns around privacy \cite{Holstein_Wortman_2019,Bogen2020,Andrus2021,patro2022fair}. %

\xhdr{Scalability in industry applications}
Web-scale ranking algorithms are required to balance recommendation performance and personalization with scalability in the number of visits, the number of items to be ranked, the amount of training data, etc. \cite{lifeed}. To this end, algorithms often target several engagement metrics at once. In the connection recommendation setting, the target utility proxy could be a mixture of models of the probability that a connection invite is send, the probability that an invite would be accepted, and measures of down-stream engagement. Multi-objective optimization provides an efficient way to derive recommendation policies which balance different business interests in real-world recommender systems \cite{clickShaping,oms}.
In our work, we compute relevance scores for connection recommendation based on a key engagement metric and intervene on fairness by re-ranking the obtained ordering.
Large-scale recommendation systems usually favor post-processing strategies like this over pre- or in-processing fairness intervention as they are typically agnostic to the underlying model structures and scale well with large amounts of data and across similar applications \cite{Geyik_Ambler_Kenthapadi_2019,Nandy_Diciccio_Venugopalan_Logan_Basu_El}.
We note that our simulation study invokes fairness criteria by directly solving separate optimization problems for each session which can lead to latency in online settings. This problem can be solved by solving an aggregate primal optimization problem and subsequently relying on a dual trick to quickly obtain re-rankings online \cite{Basu_DiCiccio_Logan_El}.

\xhdr{Homophilic behavior and algorithmic glass ceilings} 
We find that connection recommender systems can exacerbate disparities between different demographic groups of users leading to a group wise rich-get-richer effect that benefits the majority population. This finding aligns with the observations of previous research in the social networks area. \cite{Stoica2018} study the impact of gender and homophily in the context of Instagram recommendations. They authors demonstrate the existence of an algorithmic glass ceiling that prevents equal representation of women and people of color based on reinforced pre-existing disparities. 
\cite{Hofstra2017} analyze patterns of segregation based on gender and ethnicity on Facebook, and \cite{Karimi2018} focus on the effects of homophilic behavior paired with a group size difference in preferential attachment models. The study finds that smaller minority groups suffer more from homophily.
\cite{Fabbri_Bonchi_Boratto_Castillo_2020} study people recommender systems and find that homophily in demographic groups can lead to disparate visibility of minorities.
Related phenomenons have been described as the filter bubble problem of link prediction \cite{Masrour_Wilson_Yan_Tan_Esfahanian_2020,Nguyen2014} which is a term used more generally to describe how algorithmic personalization can lead to overly homogeneous recommendations essentially isolating users from different viewpoints.
Lastly, research in social psychology finds that homophilic tendencies hinder women in male dominated fields from forming professional connections which can lead to sex segregation over time \cite{Roth_2004}, and is closely related to the idea of tokenism \cite{Davis_2016,kanter1977}.%

\clearpage
\bibliographystyle{ACM-Reference-Format}
\bibliography{sample-base}

\appendix

\clearpage

\section{Supplementary figures}
\label{sec:app_fig}

\begin{minipage}{0.5\textwidth}
\centering
    \centering
    \includegraphics[trim = 20 0 10 0, clip=true, scale = 0.1]{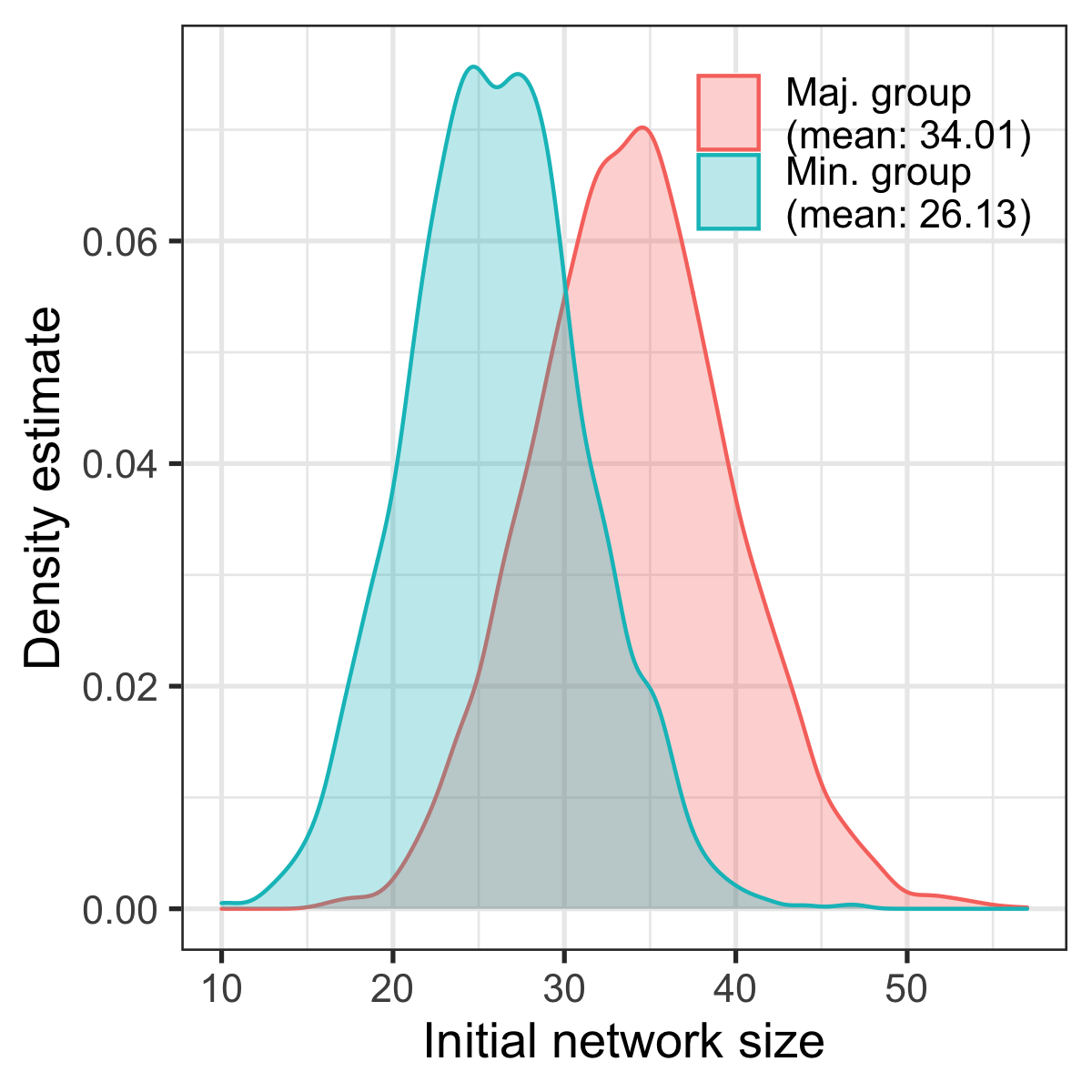}
    \includegraphics[trim = 20 0 10 0, clip=true, scale = 0.1]{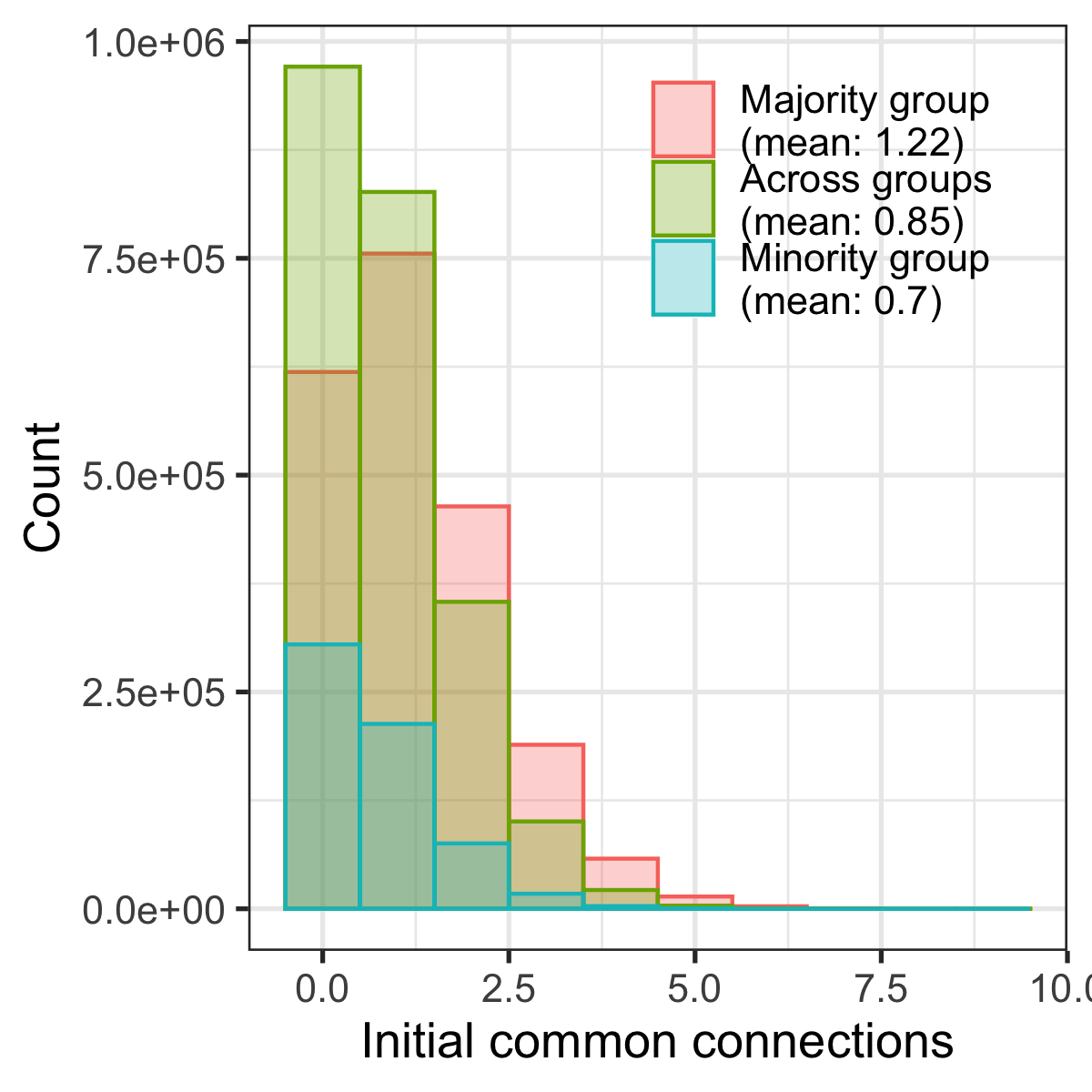}
    \includegraphics[trim = 20 0 10 0, clip=true, scale = 0.1]{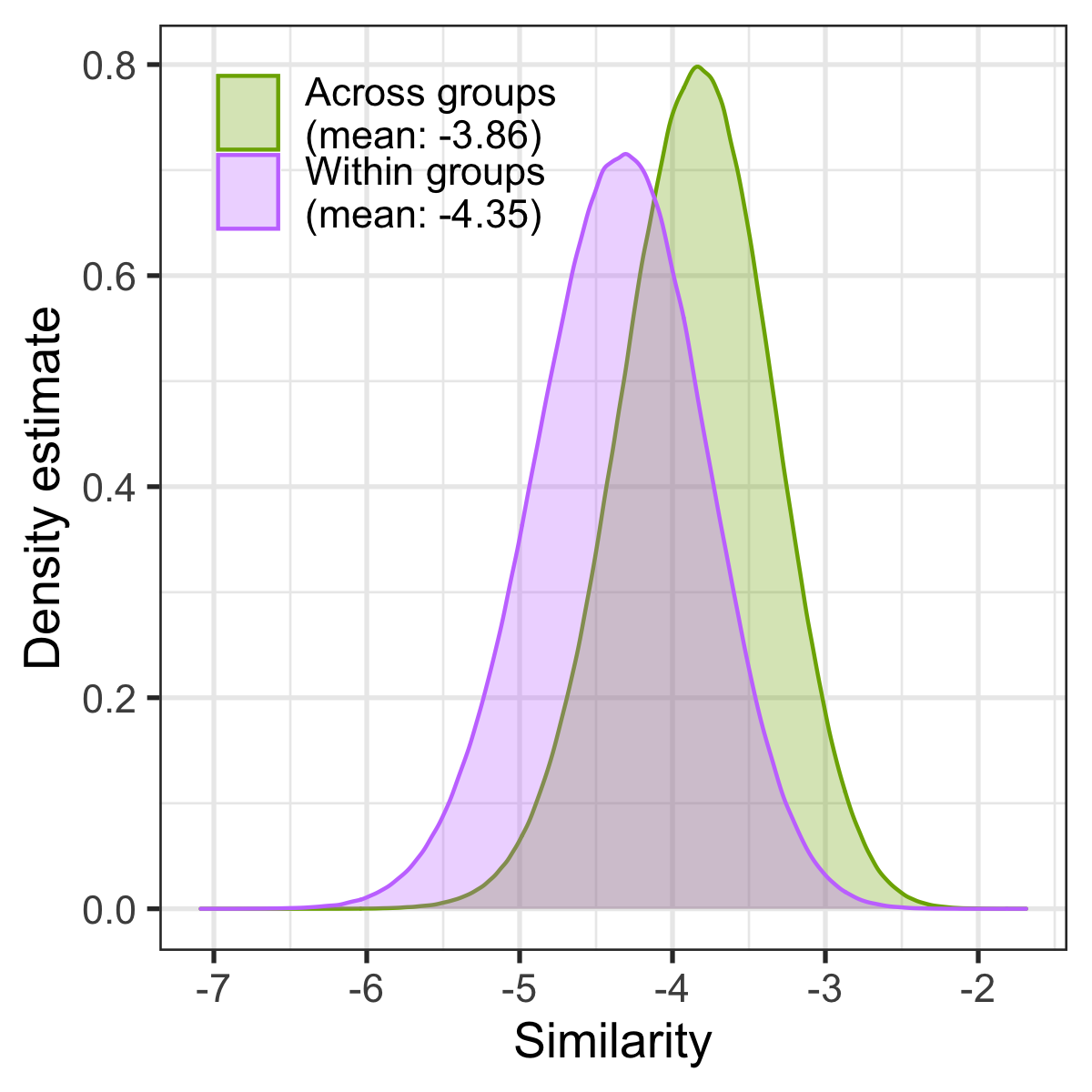}
    \includegraphics[trim = 20 0 10 0, clip=true, scale = 0.1]{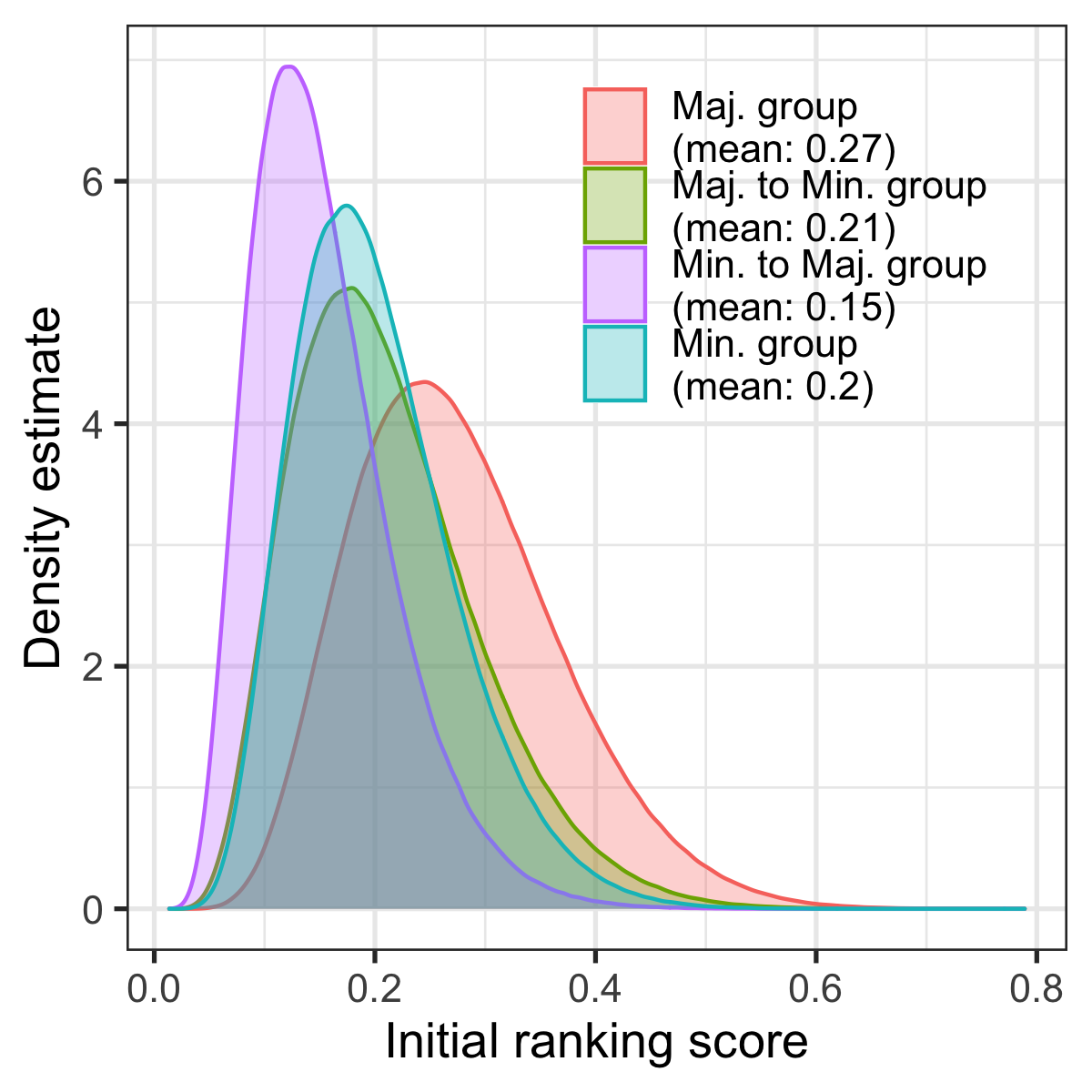}
    \captionof{figure}{
    Features of initial networks and ranking scores at $\mathbf{t=0}$ for 10 simulation runs. Networks are initialized with a stochastic block model with $\mathbf{p_\text{connect}(i,j)=0.04}$ for $\mathbf{i,j\in G_0}$ (majority group), $\mathbf{p_\text{connect}(i,j)=0.032}$ for $\mathbf{i,j\in G_1}$ (minority group) and $\mathbf{p_\text{connect}(i,j)=0.023}$ otherwise. We see that the initial network sizes tend to be larger for majority group members (top left), and members in the same groups tend to be more similar than members in different groups (bottom left). The average number of initial common connections between members in $\mathbf{G_0}$ ($\mathbf{G_1}$) is 1.22 (0.7) while the average number for member pairs across groups is 0.85 (top right). We use the scoring model to compute initial ranking scores between all unconnected members in $\mathbf{t=0}$ and see that scores tend to be higher in the majority group as compared to the minority group and for in-group pairings of members as compared to pairings across groups (bottom right).}
    \label{fig:initial}
\end{minipage}

\begin{minipage}{0.5\textwidth}
    \centering
    \includegraphics[trim = 30 0 10 0, clip=true, scale = 0.09]{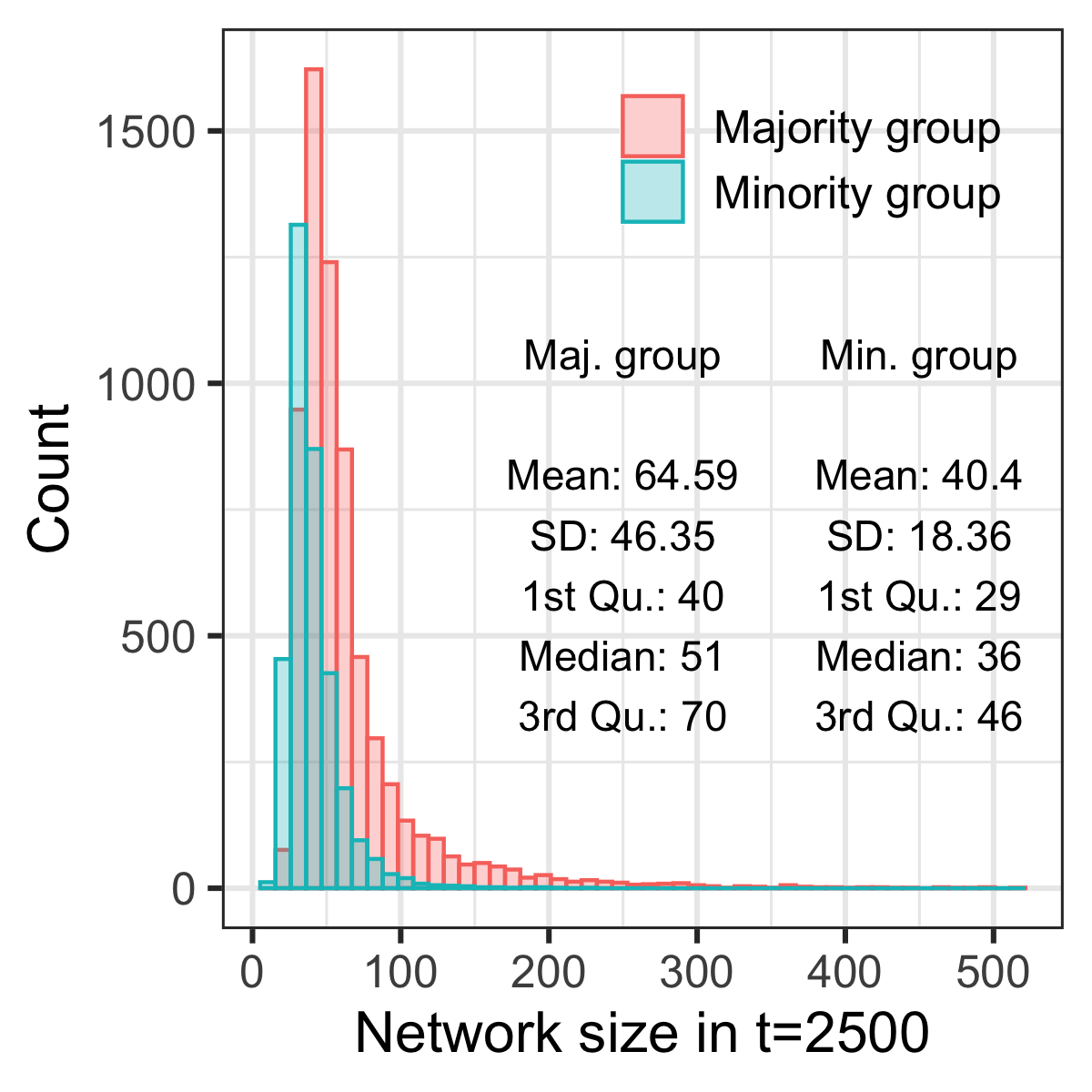}
    \includegraphics[trim = 30 0 10 0, clip=true,scale = 0.09]{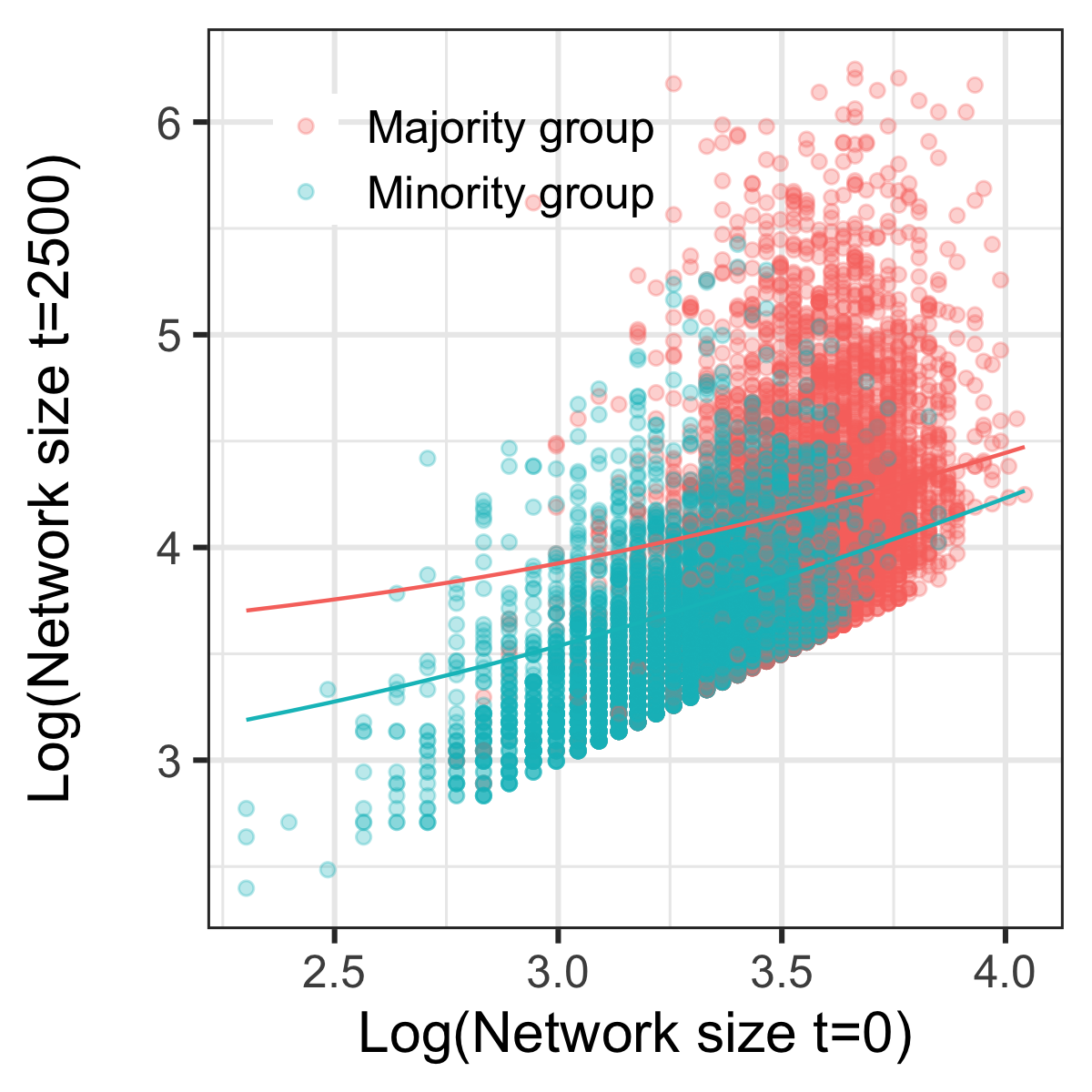}
    \captionof{figure}{Distribution of network sizes in $\mathbf{t=2500}$ without fairness intervention (left). 
    Log-log plot of network sizes without fairness intervention in $\mathbf{t=0}$ and $\mathbf{t=2500}$ with curves denoting the counterfactual network size at $\mathbf{t=2500}$ if the total increase within groups was distributed evenly (right). Results include 10 simulation runs. We see that unconstrained recommendation leads to a group-wise  rich-get-richer effect that benefits the majority group.}
    \label{fig:noReRanker}
\end{minipage}

\begin{minipage}{0.5\textwidth}
    \centering
    \includegraphics[trim = 30 0 10 0, clip=true, scale = 0.09]{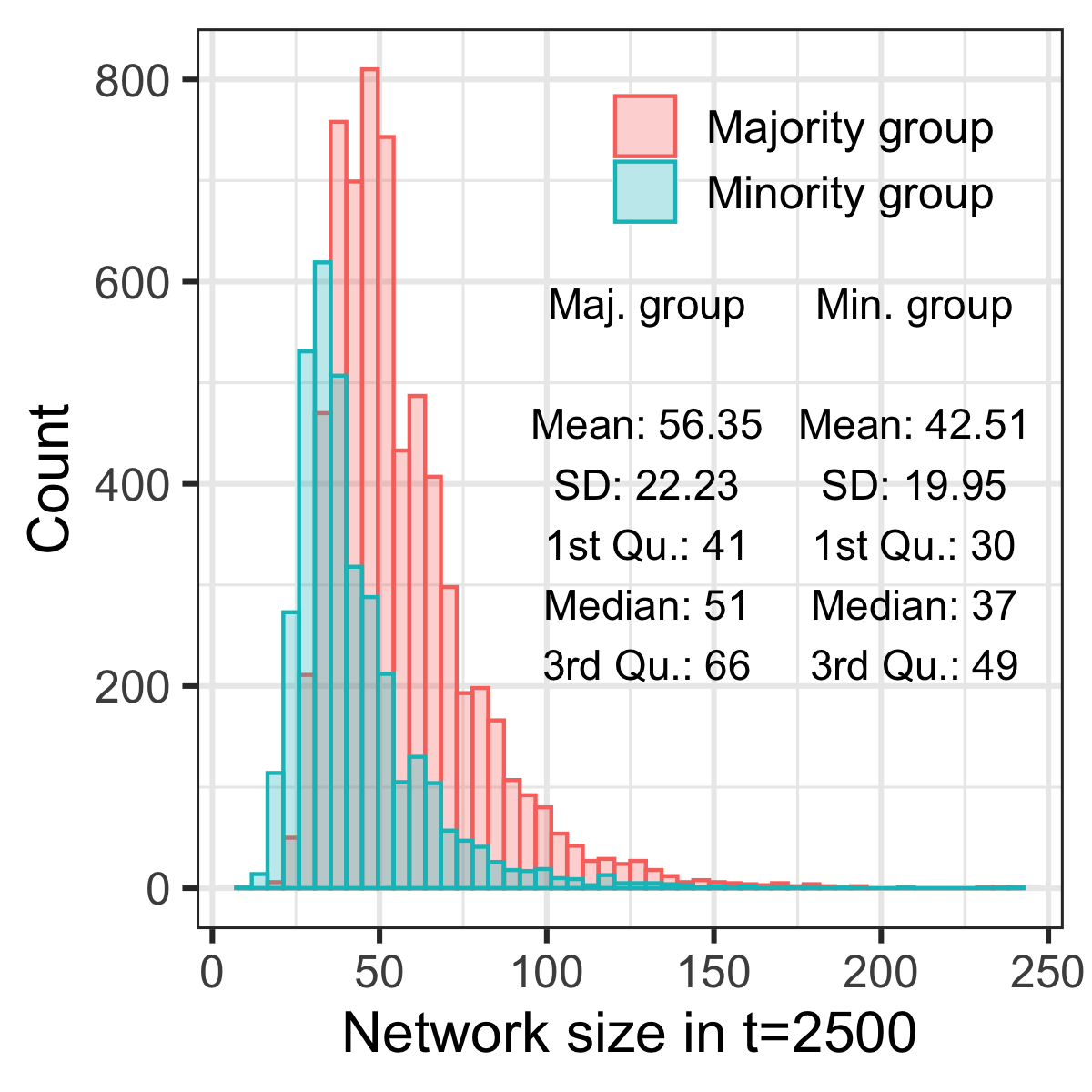}
    \includegraphics[trim = 30 0 10 0, clip=true,scale = 0.09]{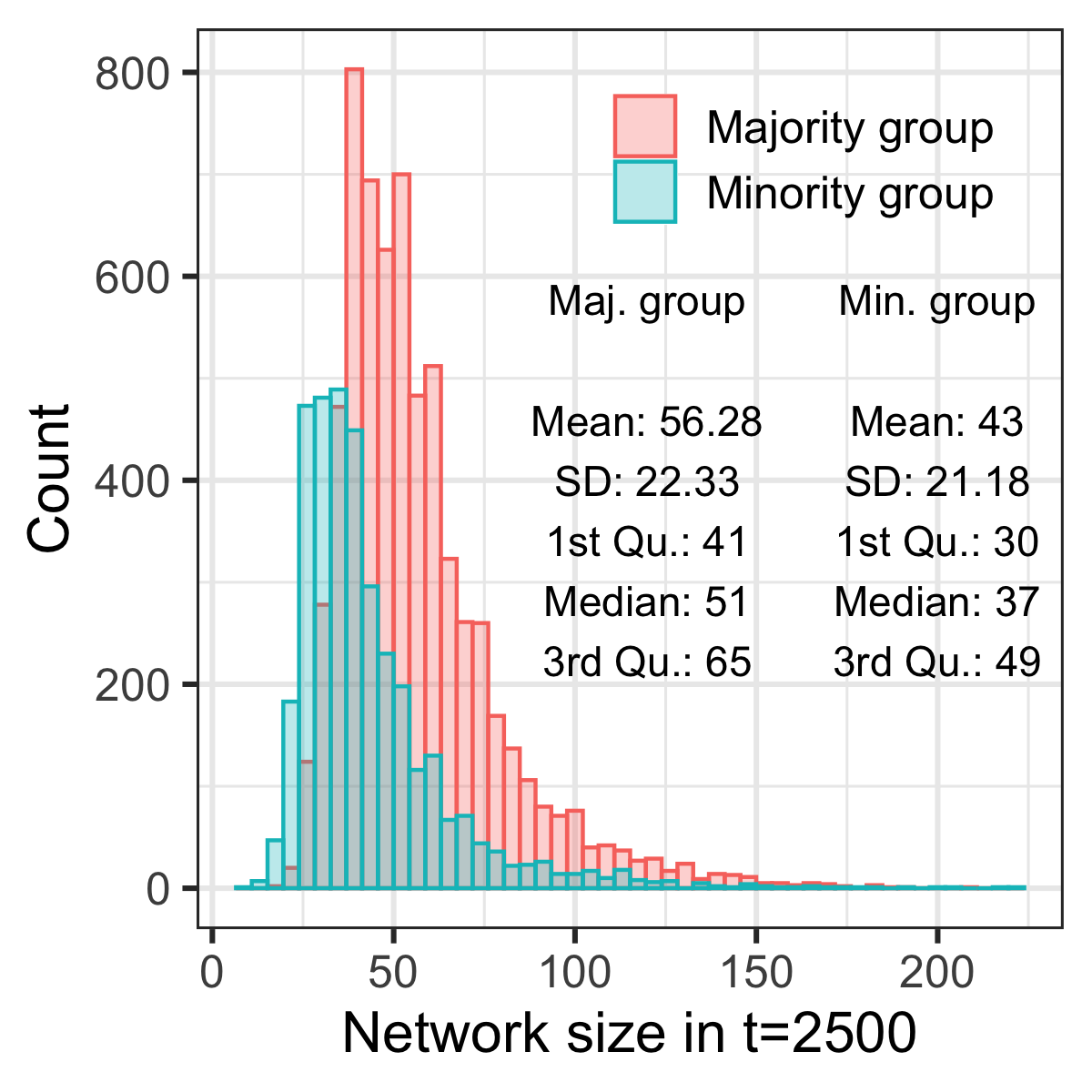}
    \captionof{figure}{Distribution of network sizes in $\mathbf{t=2500}$ after demographic parity of exposure intervention (left) and after dynamic parity of utility intervention (right). Results include 10 simulation runs. We see that both interventions are unsuccessful in aligning network size distributions across groups.}
    \label{fig:dpdyn}
\end{minipage}

\clearpage
\section{Comparison of main simulation and mixed preferential attachment model}
\label{sec:app_comparison_models}

The described Mixed Preferential Attachment (MPA) model is used to simulate a simplified mechanism of connection recommendation that qualitatively maintains many of the key aspects of the bias mechanism in our main simulation. 
First, we note that, although the main simulation assumes a fixed graph and new nodes are added at each iteration in the MPA
model, recommendations are on average sought out more frequently by the members of the majority group in both models. The simulation study models this by exponential waiting times depending on the current network size of members, while the MPA model assumes a rate $r\leq 0.5$ of minority group members.
Second, once a source member is selected, both models lead to connections to destination members based on two main features with similar interpretations.
The scoring and connection models in the main simulation make use of the similarity between group-dependent features which renders members from the same groups more likely to be suggested and connect. Meanwhile, the MPA model uses the parameters in the mixing matrix to express a similar in-group preference when $p_0,p_1> 0.5$.
In addition, ranking scores in the main simulation are positively impacted by the number of common connections between the ranked members.
This is not possible in the MPA model because source members enter the network without previous connections. However, the MPA model gives preference to destination members with large networks which goes into the same direction as the number of common connections and can lead to a similar effect \cite{LibenNowell2007}.
Overall, the models are similar enough that it is reasonable to expect that some of the qualitative observations we can make by analyzing the MPA model can be translated to insights into the behavior of the realistic simulation study on a group-aggregate level.

\section{Proof of Theorem~\ref{thm:dp}}
\label{sec:proof_dp}
Our proof follows a similar procedure to the proof of Theorem~1 in \cite{Avin_Daltrophe_Keller_Lotker_Mathieu_Peleg_Pignolet_2020}, yet in our setting we are able to obtain a relatively simple closed form solution of the limit $\alpha$.

Note that exactly one new member and one connection are added in every time step.
Given that the incoming source member is of group $G_i$, we denote the probability that the connection forms to a member of group $G_j$ by $P_{ij}$. 
We note that it holds $P_{ij}=1-P_{ii}$ for $i,j\in\{0,1\}$ and use the mixing matrix $\pi$ compute
\begin{align*}
    P_{00} &= (1-r)p_0 + (1-r)(1-p_0)P_{00} + rp_0P_{00}\\
    \Leftrightarrow P_{00} &= \frac{(1-r)p_0}{r+p_0-2rp_0},
\end{align*}
and
\begin{align*}
    P_{11} &= rp_1 + r(1-p_1)P_{11} + (1-r)p_1P_{11}\\
    \Leftrightarrow P_{11} &= \frac{rp_1}{1-r-p_1+2rp_1}.
\end{align*}
Let $N_{t+1}$ be the number of group $G_1$ degrees added in step $t+1$. Then, it holds that
\begin{align}
\begin{alignedat}{1}
\label{eq:ap1}
    \mathbb{E}[N_{t+1}] &= 2rP_{11} + rP_{10} + (1-r)P_{01}\\
    &=rP_{11} - (1-r)P_{00} + 1\\
    &=r\frac{rp_1}{1-r-p_1+2rp_1} - (1-r)\frac{(1-r)p_0}{r+p_0-2rp_0} + 1.\\
\end{alignedat}
\end{align}
We know that $\alpha_{t}=d_{t}(G_1)/d_t$ and $d_t=d_0+2t$ for all $t$.
Thus,
\begin{align*}
    \mathbb{E}[N_{t+1}]&=\mathbb{E}[d_{t+1}(G_1) - d_t(G_1)|\alpha_t]\\ &= 
    \mathbb{E}[\alpha_{t+1}\vert \alpha_t]d_{t+1}-\alpha_{t}d_{t}\\ &= \mathbb{E}[\alpha_{t+1}\vert \alpha_t](d_0 + 2(t+1))-\alpha_{t}(d_0 + 2t),
\end{align*}
and with we receive
\begin{align*}
    \mathbb{E}[\alpha_{t+1}\vert\alpha_t] = \frac{\alpha_t (d_0 + 2t) + \mathbb{E}[N_t]}{d_0 + 2(t+1)} = \alpha_t + \frac{\mathbb{E}[N_t] - 2\alpha_t}{d_0 + 2(t+1)}.
\end{align*}
Recursively inserting the conditional expected values of $\alpha_i$ for $i\in[t]$ and shifting $t$ by one gives
\begin{align*}
    \mathbb{E}[\alpha_{t}] &= \alpha_0 \prod_{j=1}^t \left(1-\frac{2}{d_0+2j}\right) + \sum_{i=1}^t \left(\frac{\mathbb{E}[N_t]}{d_0+2i}\prod_{k=i+1}^t \left(1-\frac{2}{d_0+2k}\right)\right)\\
    &= \alpha_0 \prod_{j=1}^t \left(1-\frac{2}{d_0+2j}\right) + t\frac{\mathbb{E}[N_t]}{d_0+2t}.
\end{align*}
Note that
$$
\lim_{t\to\infty}\prod_{j=1}^t \left(1-\frac{2}{d_0+2j}\right) = 0
$$
and thus with Equation~\eqref{eq:ap1}
\begin{align*}
\lim_{t\to\infty}\mathbb{E}[\alpha_t] &= \lim_{t\to\infty}\frac{\mathbb{E}[N_t]}{d_0/t+2}\\ &= \frac{\mathbb{E}[N_t]}{2}\\ &= 
\frac{1}{2}\left(r\frac{rp_1}{1-r-p_1+2rp_1} - (1-r)\frac{(1-r)p_0}{r+p_0-2rp_0} + 1\right).
\end{align*}

\section{Proof of Theorem~\ref{thm:dynamicintervention}}
\label{sec:proof_dyn}

We use the same notation as in Appendix~\ref{sec:proof_dp} and note that in order for the limiting share $\alpha$ to be independent of $p_0$ and $p_1$, the same must be true for the conditional probabilities $P_{ij}$.
Instead, the $P_{ij}$ must be chosen such that the expected number of $G_1$ balls added in $t$ fulfills
$$
    \mathbb{E}[N_{t}] = rP_{11} - (1-r)P_{00} + 1 = 2r, 
$$
since $\lim_{t\to\infty}\mathbb{E}[\alpha_t]=\mathbb{E}[N_t]/2$ (see proof of Theorem~\ref{thm:dp}).
This is trivially fulfilled by assuming $P_{00}=P_{11}=1$ as described in Section~\ref{sec:dyn}. A more interesting solution is obtained by crafting an intervention which ensures that $P_{11}=r$ and $P_{00}=1-r$ which we will do in the following. 

We recompute the probabilities $P_{ij}$ as functions of the mixing matrix $\pi$ and the rejection sampling probabilities $q_{ij}$.
In iteration step $t+1$, we have
\begin{align*}
    P_{00}=&\ (\alpha_t-1)q_{00}p_0 + (\alpha_t-1)q_{00}(1-p_0)P_{00}\\ &+ 
    (\alpha_t-1)(1-q_{00})P_{00} + \alpha_t q_{01}(1-p_0)P_{00} + \alpha_t (1-q_{01})P_{00}.
\end{align*}
Assume that $q_{ij}=1-q_{ii}$ for $i\neq j$.
Setting $P_{00}=1-r$ and solving for $q_{00}$, yields that we need to set 
$$
    q_{00}=\frac{(1-r)(\alpha_t(p_0-2)+2)}{p_0(\alpha_t-r)}
$$
as long as $\alpha_t\neq r$. Note that $p_0,p_1$ and $\alpha_t$ are bounded away from 0 and 1. Similarly, it holds that
\begin{align*}
    P_{11}=&\ \alpha_tq_{11}p_1+\alpha_tq_{11}(1-p_1)P_{11}+\alpha_t(1-q_{11})P_{11}\\ &+(1-\alpha_t)q_{10}p_1P_{11}+(1-\alpha_t)(1-q_{10})P_{11}.
\end{align*}
We set $P_{11}=r$ and receive
$$
q_{11}=\frac{(1-\alpha_t)(1-p_1)r}{\alpha_t(p_1-r)-p_1r+r}
$$
for $\alpha_t(p_1-r)-p_1r+r\neq 0$ and the claim follows.

\end{document}